\documentclass{aastex}
\usepackage{emulateapj5}

\begin{document}

\slugcomment{To appear in The Astrophysical Journal}

\title{Molecular Tracers of the Central 12 pc of the Galactic Center}

\author {Melvyn C. H. Wright\altaffilmark{1}, Alison L. Coil\altaffilmark{1}, Robeson S. McGary\altaffilmark{2}, Paul T.P. Ho\altaffilmark{2} and Andrew.I.Harris\altaffilmark{3}}
\altaffiltext{1}{Radio Astronomy Laboratory, University of California, Berkeley, CA 94720 E-mail: mwright@astro.berkeley.edu, acoil@astro.berkeley.edu}
\altaffiltext{2}{Harvard-Smithsonian Center for Astrophysics, 60 Garden St., Cambridge, MA 02138 E-mail: rmcgary@cfa.harvard.edu, ho@cfa.harvard.edu}
\altaffiltext{3}{University of Maryland, E-mail: harris@astro.umd.edu}

\begin{abstract}
 
We have used the BIMA array to image the Galactic Center with a
19-pointing mosaic in HCN(1-0), HCO$^+$(1-0), and H 42$\alpha$ emission
with 5 km s$^{-1}$ velocity resolution and $13'' \times 4''$ angular
resolution.  The $5'$ field includes the circumnuclear ring (CND) and
parts of the 20 and 50 km s$^{-1}$ clouds.  HCN(1-0) and HCO$^+$ trace the
CND and nearby giant molecular clouds while the H 42$\alpha$ emission
traces the ionized gas in Sgr A West.  We find that the CND has a
definite outer edge in HCN and HCO$^+$ emission at $\sim45''$ radius and
appears to be composed of two or three distinct streams of molecular
gas rotating around the nucleus.

Outside the CND, HCN and HCO$^+$ trace dense clumps of high-velocity gas in 
addition to optically thick emission from the 20 and 50 km s$^{-1}$ clouds.  A
molecular ridge of compressed gas and dust, traced in NH$_3$ emission
and self-absorbed HCN and HCO$^+$, wraps around the eastern edge of Sgr A
East.  Just inside this ridge are several arcs of gas which have been
accelerated by the impact of Sgr A East with the 50 km s$^{-1}$ cloud.

HCN and HCO$^+$ emission trace the extension of the northern arm of Sgr A
West which appears to be an independent stream of neutral and ionized
gas and dust originating outside the CND. Broad line widths and OH
maser emission mark the intersection of the northern arm and the CND.

Comparison to previous NH$_3$ and 1.2mm dust observations shows that
HCN and HCO$^+$ preferentially trace the CND and are weaker tracers of
the GMCs than NH$_3$ and dust.  We discuss possible scenarios for the
emission mechanisms and environment at the Galactic center which
could explain the differences in these images.

\end{abstract}


\section{Introduction}

A 2-5 pc circumnuclear ring (CND) of neutral gas and dust has been 
imaged at the Galactic Center in the far-infrared and radio
\citep{gen85,gus87,mar93}.  It has been proposed that the CND
is part of an accreting disk of material surrounding the massive
central source Sgr A$^*$ \citep{gus87}, an apparent supermassive black hole
\citep{ghe98,gen97}.  Ionized gas streamers appear to be infalling
towards the central mass, originating at the inner edge of the
CND and converging at the location of the bar near Sgr A$^*$
\citep{loe83}.  It is unclear whether the CND is a transitory or
stable structure.  If it is stable, the question remains as to how the
CND itself is maintained.  

At a projected distance of 10 pc from the Galactic Center lie two
giant molecular clouds (GMC), the 20 km s$^{-1}$ cloud
(M-0.13-0.08) and the 50 km s$^{-1}$ cloud (M-0.02-0.07)
\citep{gus81} to the south and east of the CND respectively.
\citet{gus81b} argued that these clouds must lie more than 100 pc from
the Galactic Center in order not to be tidally disrupted. Their
interpretation of formaldehyde absorption data places the 50 km
s$^{-1}$ cloud behind the nucleus and the 20 km s$^{-1}$ cloud in front.
Various studies have investigated the interactions
between these molecular clouds, the non-thermal radio source Sgr A
East, the CND and the ionized gas streamers in the nucleus
\citep{hoe85, gen90, hoe91, ser92, den93, mar95, coi00}.  There are
suggestions of molecular gas steamers from both of the GMCs located 10
pc away in projection that may be feeding the CND.  A southern
streamer has been mapped in NH$_3$(1,1), (2,2) and (3,3) and
sub-millimeter continuum \citep{hoe91, den93, coi99} and emission in
the same region has been seen in HCN(3--2) \citep{mar95}.  An eastern
streamer, lying at the edge of the Sgr A East SNR and protruding in
towards the nuclear region, has been mapped in HCN(1--0) \citep{hoe93}
and HCN(4--3) and HCN(3--2) \citep{mar95}.  This streamer has not been
seen in NH$_3$ or sub-millimeter continuum.  More recently,
\citet{coi00} presented images of two streams of NH$_3$(1,1) and (2,2)
emission parallel to the Galactic plane, which they interpret as
evidence for infall from the 20 km s$^{-1}$ cloud onto the CND, and as
evidence for interactions between Sgr A East and the 50 km s$^{-1}$
cloud, and a SNR to the south of Sgr A East (G359.92--0.09). They
argue that both clouds are within 10 pc of the nucleus and that only
the 20 km s$^{-1}$ cloud is interacting with the CND.

Many of these observations have a limited field-of-view and velocity
coverage, mapping only the central $2-4'$ (5--10pc) of the Galaxy and
not extending into the large molecular clouds, thereby unable to
thoroughly address the question of possible interactions with the
circumnuclear region.  This paper describes observations with the BIMA
array\footnote{The BIMA array is operated by the
Berkeley-Illinois-Maryland Association under funding from the National
Science Foundation.} of the central $5'$ in HCO$^+$, HCN(1-0), and H
42$\alpha$ emission with 5 km s$^{-1}$ resolution, and in 87 GHz
continuum emission.  The $5'$ field of view corresponds to a field
roughly 12.5 pc at the distance of the Galactic Center (assuming
$R_\odot$ = 8.5 kpc).

\section{Observations}

\begin{figure}
\centering
\includegraphics[angle=270,width=6in]{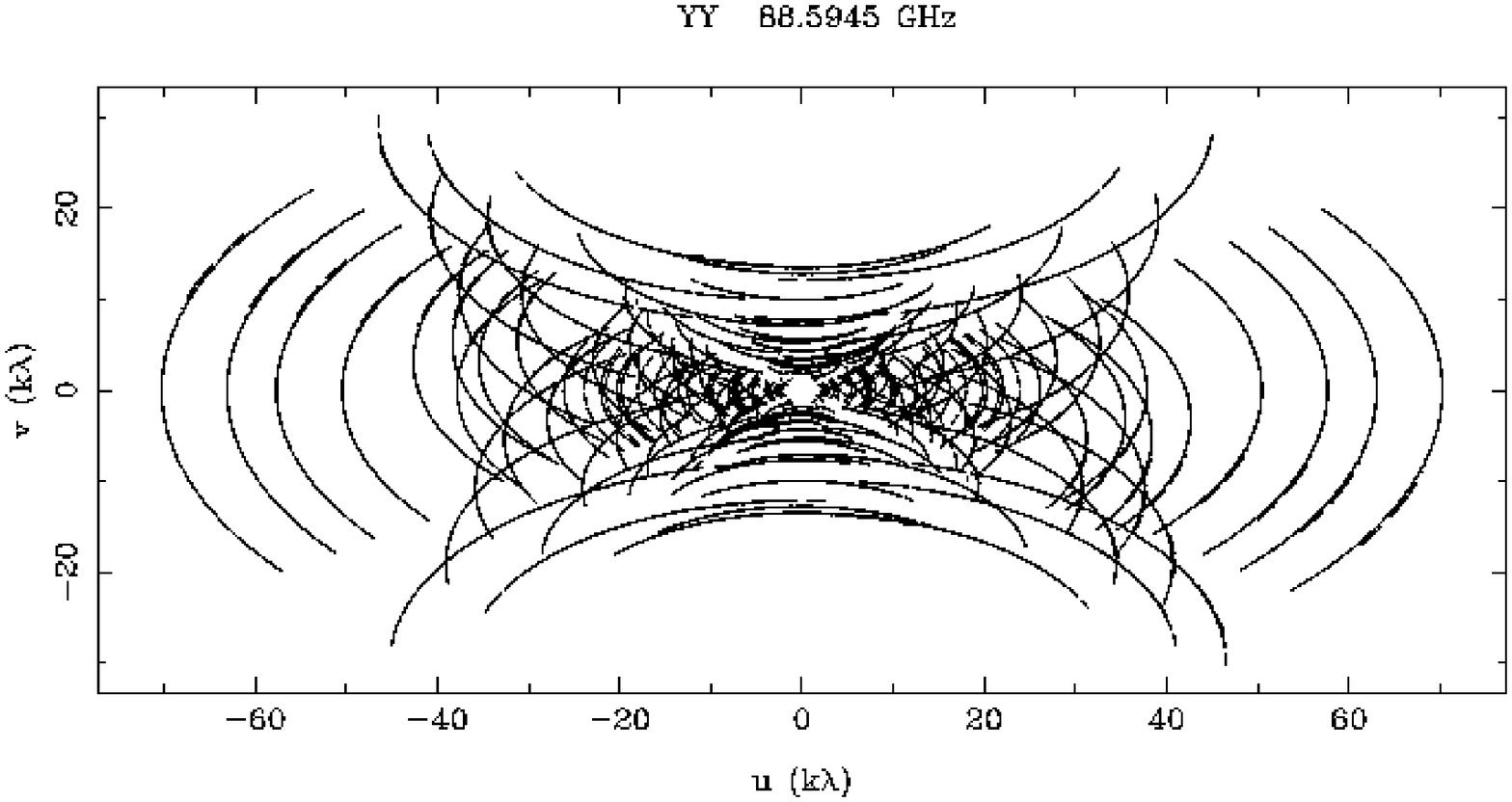}
\caption{$uv$ coverage of our BIMA observations.} \label{uvcoverage}
\end{figure}

Observations were obtained with the BIMA array \citep{wel96}
between 1998 March and 1999 October using four antenna configurations with
antenna separations from 6 to 240m, including the recently completed,
compact D--array which gives good $uv$ coverage down to the 6.1 m
antenna diameter, and a new C--array which greatly improves the $uv$
coverage for low declination sources.  Figure \ref{uvcoverage} shows the 
$uv$ coverage. The resulting synthesised beam width (FWHM) 
is $13.2'' \times 4.3''$.

We mapped the Galactic Center simultaneously in HCN (J=1-0), HCO$^+$
(J=1-0), and H 42$\alpha$ emission with 5 km s$^{-1}$ resolution over
a velocity range of 300 km s$^{-1}$ centered close to v$_{LSR}$ = 0 km
s$^{-1}$ for all three spectral lines.  We also mapped 87 GHz
continuum emission with 400 MHz bandwidth in each sideband of the
local oscillator.  We used a 19-pointing hexagonal mosaic with $1'$
spacing centered on Sgr A*.  Additional observations were obtained in
a single pointing centered on Sgr A* to enhance the sensitivity to
emission within the central $2'$ and to compare with the images
obtained from the mosaic observations.

We observed the quasar 1733-130 as a phase calibrator at 30 to 35 min
intervals after 2 complete cycles of 19 pointings with 35 s
observations at each pointing. This cycle of observations was repeated
for 4 to 5 hours while the Galactic Center was above $\sim$12$^\circ$
elevation. Uranus was observed to check the flux density
calibration, and the strong quasar 3C273 was used for the bandpass 
calibration. The flux density scale is accurate to about 10\%,
and the on-line bandpass calibration is good to about 5\%. The data
were calibrated and imaged using the MIRIAD software package \citep{sau95}.

\section{Image Fidelity}

In order to estimate the errors in our images, we processed the data in
a number of different ways and compared the resulting images.  Images
using different subsets and weightings of the $uv$ data are consistent
to about 5\% RMS of the peak value.

Single sideband system temperatures during the observations ranged from
170 to 1600 K scaled to outside the atmosphere.  Images were made by
weighting the $uv$ data by the inverse square of the system
temperature.  Robust weighting was used as a compromise between natural
weighting, which gives the lowest thermal noise, and uniform weighting
of the $uv$ plane.

We deconvolved the images using both the SDI CLEAN algorithm
\citep{ste84}, and the maximum entropy algorithm (MEM) \citep{sau96}.
To avoid bias in the MEM images, we used an entropy measure
-log(cosh(p)) -- also called the maximum emptyness criteria, where p is
the pixel value. This entropy measure handles the complex regions of
emission and absorption seen in the spectral line images.
Images of individual pointings are consistent with the mosaic images.

The noise level varies with radius in the mosaic images.  For the
spectral line images, the thermal noise level is 0.07 K in 20 km
s$^{-1}$ channels at the map center, increasing to 0.11 K at $120''$
radius and 0.15 K at $150''$ radius.
We compared spectral line images obtained by interchanging the order of
velocity averaging, deconvolution, and continuum subtraction.
The 20 km s$^{-1}$ images are consistent with an RMS difference over
the central $5'$ field of 0.1-0.2 K, and a
maximum difference (between $120''$ and $150''$ radius) of 0.7 K.
At the 5 km s$^{-1}$ resolution in the spectra which follow, the
measured rms noise level is around 0.5 K in a $13.2'' \times 4.3''$
synthesised beam, giving a dynamic range of about 20:1.
We also compared images of the 85 and 89 GHz data in each sideband of the
local oscillator. The rms difference between these images is 35 mJy
with a maximum difference of 100 mJy. The theoretical thermal noise in
the difference image is 25 mJy.

\subsection{Data Sampling}

The hexagonal pointing mosaic with $1'$ spacing is well sampled at 89
GHz (with a primary beam FWHM of $2.15'$), giving images with
approximately uniform sensitivity across a $5'$ field.  For a mosaic
with 19 pointing positions, the Nyquist sample interval is 9 s on the
maximum 240 m baseline used in these observations. Our 35 s sample
interval was chosen to minimize the overall noise in the images, taking
into account the observing efficiency and the data sampling.  Repeated
observations of each $uv$ track (Figure \ref{uvcoverage}) extend the
Nyquist sampling of the data along the $uv$ tracks to 100 m (30
k$\lambda$) baselines, as well as fill in many calibration gaps.
Compact structures are adequately sampled out to the maximum 240 m (75
k$\lambda$) baseline.  Extensive modeling of mosaic images with
undersampled data \citep{wri99} shows a slow degradation of image
fidelity with decreasing sampling.

\subsection{Missing Short Spacings}

These images do not include any observations shorter than the minimum
interferometer spacing of 6.1 m. The missing short spacings contain
information about structures larger than about $1'$.  Both SDI and MEM
deconvolutions are very similar if we use an entropy measure
-log(cosh(p)).  For the HCN and HCO$^+$ emission, which are heavily
absorbed by foreground gas, the spatial filtering provided by the
interferometer is an advantage in isolating warmer material associated
with the CND, but resolves out the large scale structure of the nearby
molecular clouds.

In summary, structures with angular scales between $1'$ and the $13''$
x $4''$ resolution are well sampled in these images, giving an image
fidelity close to the thermal noise, or 5\% of the peak value in each
spectral channel, whichever is greater.

\section{Results}

\subsection{Continuum Emission}

\begin{figure}
\centering
\includegraphics[angle=270,width=6in]{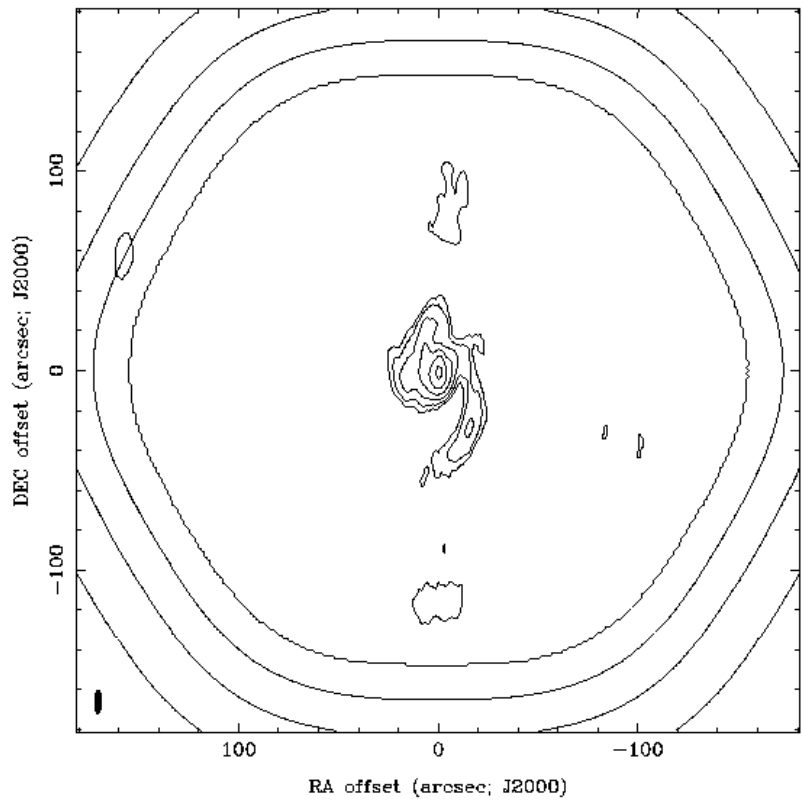}
\caption{Continuum emission at contour levels
.05,.1,.2,.4,.8,1.6 Jy beam$^{-1}$.
The synthesised beam FWHM is shown in the bottom left corner.  The
effective primary beam attenuation in the mosaic image is indicated at
levels 0.25, 0.5, 0.75, 1.0.} \label{cont}
\end{figure}

Figure \ref{cont} shows 87 GHz continuum emission over an 800 MHz
bandwidth averaged over line-free spectral windows in both sidebands of
the local oscillator.  We imaged the eight 100 MHz continuum bands in
both sidebands of the local oscillator and found that the images were
consistent.  We then averaged the upper and lower sidebands
separately.  These two images agree within an 8\% scale factor and 14
mJy rms, consistent with calibration errors (10\%) and thermal noise
(13 mJy/beam) respectively. The upper and lower sidebands were combined
to produce the image shown in Figure \ref{cont}. In this and the
following mosaic images, the primary beam attenuation is fully
corrected out to $150''$ radius, falling to 0.5 at $180''$.  This
attenuation can be seen in the contours in Figure \ref{cont}.  The rms
noise level in the averaged continuum image is 6 mJy/beam at the
center, increasing to 13 mJy/beam at $150''$ radius.  Within the $5'$
field we can see the thermal mini-spiral, the central bar, the western
arc curving around to the southwest, and the non-thermal emission from
Sgr A* (see Lacy et al. 1991 Figure 1 for a discussion of features in
the mini-spiral).  The features at $\sim100''$ north and south are
alias responses resulting from the discrete $uv$ sampling.

The SDI and MEM deconvolutions result in an integrated flux density out
to $30''$ of 8.9 Jy and 14.2 Jy, respectively.  The integrated flux
density estimated from the shortest $uv$ spacings (6.1 m) is 15$\pm$1
Jy, with considerable structure apparent on arcmin scales. The MEM
deconvolution has recovered most of the flux density observed on the
shortest interferometer spacings, but falls short of the $\sim$ 27 Jy
obtained with single dish observations \citep{wri93}. The peak flux
density is 2.14 Jy/beam and includes emission from both Sgr A* and the
region of the bar a few arcseconds to the south. The average flux
density of Sgr A* itself, estimated from both the $uv$ data and the
image deconvolution, is 1.7 to 1.8 Jy/beam, which is consistent with
previous observations of Sgr A* at this resolution \citep{wri93}.
Although the flux density of Sgr A* may have varied during the
observations, the present data do not include sufficiently long $uv$
spacings to reliably estimate this.

Gaussian fits to the compact continuum emission ($180''$ east, 
$60''$ north) give a peak flux density of 160 mJy/beam and an
integrated flux density of 510 mJy with an overall error of $\sim$10\%
(after correction for the residual primary beam attenuation).  The
fitted position is $\alpha_{2000}$=17$^h$ 45$^m$ 52$^s$.05,
$\delta_{2000}$=--28$^\circ $59$^m$ 30$^s$.91, consistent with a blend
of two compact sources, Sgr A-A and Sgr A-B, with a total flux density
of 337 mJy at $\lambda$=6 cm \citep{hoe85}. These sources are in a
string of compact HII regions with shell-like structures located to the
east of the Sgr A East shell \citep{eke83,gos85,yus87}.

\subsection{H 42$\alpha$ Emission}

Figure \ref{h42} shows the H 42$\alpha$ emission integrated from --150
to +150 km s$^{-1}$. This image was formed by first subtracting the
lower sideband continuum image from the H 42$\alpha$ spectral channels
averaged in 20 km s$^{-1}$ intervals, and then averaging the SDI
deconvolution of these images.  The H 42$\alpha$ emission traces the
same thermal continuum emission as Figure \ref{cont}.  The western arc
is clearly resolved, but the northern arm and bar are blended together
at this angular resolution.  Figure \ref{h42spec} shows the H
42$\alpha$ spectra at 20 km s$^{-1}$ resolution for the positions
marked in Figure \ref{h42}.  H 42$\alpha$ emission extends over the
full velocity range sampled.  Emission at high positive velocities is
seen in the central bar close to Sgr A* (spectrum A), while the
northern arm is prominent from +50 to +150 km s$^{-1}$, the upper edge
of the spectral window (spectrum B).  Emission from the western arc
extends from --50 to --150 km s$^{-1}$, the lower edge of the window
(spectra C and D).  Our H 42$\alpha$ image is consistent with the H
92$\alpha$ emission described in detail by \citet{rob93}.

\subsection{Molecular Emission}

Figure \ref{hcn20} shows the deconvolved images of HCN(1-0) emission
in 20 km s$^{-1}$ intervals, made by subtracting the upper sideband
continuum image from 20 km s$^{-1}$ averages of HCN(1-0) spectral
channels and then deconvolving these 20 km s$^{-1}$ images.  The
rotating circumnuclear ring (CND) can be seen at a radius of
$\sim$$30''$ to $60''$ in many channels and is most obvious as the
arc--like structures $40''$ south at a velocity of --70 km s$^{-1}$
and $40''$ north at velocities of 70 and 90 km s$^{-1}$.  Features
associated with nearby GMCs are also
visible. At 30 km s$^{-1}$, emission at the southern edge of the image
(--100$''$ to --140$''$) is associated with the 20 km s$^{-1}$
cloud.  A north-south ridge of emission at 30-90 km s$^{-1}$,
located $120''$ to the east, traces gas compressed by the expansion of
Sgr A East into the 50 km s$^{-1}$ cloud \citep{ser92}.

HCO$^+$ emission in 20 km s$^{-1}$ intervals, made in the same way as the
HCN(1-0) images, is shown in Figure \ref{hco20}.  The HCO$^+$ and
HCN(1-0) emission trace the same features, though the HCO$^+$ emission is
generally weaker.  A prominent feature in HCN(1-0) which is {\it not}
seen in HCO$^+$ is the ``70 km s$^{-1}$ cloud'' apparent from 30 km
s$^{-1}$ to 70 km s$^{-1}$ on the western edge of the CND.

Figure \ref{hcn-hco} presents a comparison of HCN(1-0) and HCO$^+$ in the
central 15 pc of the Galaxy.  Contours of velocity-integrated
HCN(1-0) emission spanning --150 km s$^{-1}$ to +150 km s$^{-1}$ are
superposed on the background of HCO$^+$ emission in the same velocity
range.  The integrated HCN emission is essentially the same as the
HCO$^+$, except for deeper absorption features in the HCO$^+$.  The CND is
seen as an almost complete ring with two opposing lobes of strong
emission, surrounded by lower-level, clumpy emission.

Our results are fully consistent with earlier observations of the CND
at similar resolution \citep{gus87, mar93}. Emission from the CND is
from warm (T $\sim$150 - 450 K), clumpy gas with molecular density
$\sim$10$^5$ - 10$^6$ cm$^{-3}$ \citep{gen85}.  
HCN and HCO$^+$(1-0) are both optically thick and have
similar excitation requirements and should have similar distributions
apart from chemistry, shock and ionization effects.  While both
HCN(1-0) and HCO$^+$ trace gas with a molecular density $\sim$10$^5$ -
10$^6$ cm$^{-3}$, HCO$^+$ has a larger collision cross section and is
subject to greater absorption along the line of sight.

Figure \ref{specpos} marks positions of HCN(1-0) and HCO$^+$ spectra
shown in Figures \ref{hcnspec} and \ref{hcospec}.  Spectrum A, at the
location of Sgr A*, shows several absorption features. The narrow
features at --30, --50, and --135 km s$^{-1}$ are thought to be due to
galactic rings or arms at radii of 4 kpc (Menon \& Ciotti 1970), 3 kpc
(Oort 1977), and 250 pc (Scoville 1972; Listz \& Burton 1978; Bieging
etal 1981). The absorption from --10 to +70 km s$^{-1}$ suggests that
the 20, 50 and 70 km s$^{-1}$ gas clouds in the Galactic Center region
extend in front of Sgr A*.

Several other spectra not near Sgr A* also show absorption dips
(e.g. R, S, T, U, V \& X).  These absorption features may be due to
intervening gas or self-absorption.  However, negative features also
appear in the spectra as a result of undersampling of the data in the
$uv$ plane.  As seen in Figures \ref{hcn20} \& \ref{hco20}, there are
artificial negative features surrounding structures on scales greater
than 60$''$, particularly at velocities from 10 to 90 km s$^{-1}$
where emission from the extended molecular clouds is partially
resolved.  However narrow self-absorption features corresponding to
the 20, 50 and 70 km s$^{-1}$ clouds can be identified in
many of the HCN and HCO$^+$ spectra.

\section{Discussion}

The origin of the CND is unclear.  The central 2 arcminutes may have been
swept clear of molecular material by a wind from the central black hole
or correlated supernovae explosions (see a recent review by Morris \&
Serabyn 1996 and references therein).  The CND may also be fed from
outside the nucleus with gas from nearby GMCs.  It is also not
understood whether the CND is a long-lived or a transient feature.  As
seen in Figures \ref{hcn-hco} and \ref{specpos} as well as in
\citet{gen85}, \citet{gus87} and \citet{mar93} the CND is not a complete ring and lacks
emission in the southeast. This gap could be due to intervening gas
along the line of sight or could reflect an intrinsic lack of material
in the region.  If the CND consists of several distinct streamers in
general rotation around the nucleus, it is possible that the gas fell
towards the center from farther away, creating a disk or ring structure
which is not necessarily a stable or long-term feature. These
observations probe the spatial extent and kinematics
of the central 5$'$ surrounding the nucleus.  By determining if the CND
has a well-defined sharp outer edge, we can limit theories on its
formation.  In addition, the kinematic information can show if the CND
is a coherent feature or composed of distinct clouds.

\subsection{Spatial Extent of the CND}

HCN and HCO$^+$ emission is seen across the entire field of Figure
\ref{hcn-hco}, however the CND itself stands out clearly in the
center.  The inner edge of the CND is well-delineated by the central
cavity surrounding Sgr A* and Sgr A West.  The outer edge of the CND
is less sharply defined than in earlier single-pointing images
\citep{gen85,gus87,mar93} which are attenuated by the primary beam
cutoff, but there still appears to be an outer edge.  Figure
\ref{radprofile} shows a radial profile of HCN(1-0) emission averaged
around elliptical annuli. Following the steep rise from the center,
there is a bright narrow ring peaked at $\sim45''$ radius.  The
average brightness of the emission quickly drops from 45$''$ to 80$''$
indicating that the CND is a ring rather than a disk.  

CO studies (Lugten et al. 1986, Harris et al. 1985) of this region
show emission out to 140$''$ to the northeast and southwest, along the
major axis of the ring, which appears to follow the general rotation
pattern of the CND.  In Figures \ref{hcn20} \& \ref{hco20}, images
from --130 to --110 km sec$^{-1}$ show emission from the southwest
lobe of the CND extending to 120$''$.  Position velocity cuts $b$, $c$
and $e$ also show high-velocity HCN emission at $\sim100''$ to the
southwest of the CND.  This emission has broad linewidths ($\sim60$ km
s$^{-1}$) and may represent either a continuation of the CND or a high-
velocity streamer in this direction.  However, there is no comparable
extension of the northeast lobe.  There is instead gas at +50 to +90
km sec$^{-1}$ associated with the expansion of Sgr A East, which has
swept up and compressed gas in the 50 km sec$^{-1}$ cloud into a
distinct streamer located 150$''$ from the nucleus.  This asymmetry
along the major axis may account for some of the earlier
disagreements in the apparent rotation curve derived from lower
resolution observations (e.g. see figure 7 in \citet{gus87}).
 
In all other directions, there is a noticeable {\it lack} of emission
outside 50$''$, especially along the eastern edge between 60$''$ and
80$''$ radius.  Far from the CND, weak filamentary structures dominate
the HCN and HCO$^+$ emission. 
As discussed below in section 5.5, absorption
effects are important in the GMC's,  so that the fall-off in the
radial distribution is not as severe as shown in figure 11 in these
directions. To the west, the ``70km s$^{-1}$ cloud''
does not fit the rotation pattern, and $100''$ to the east there are
large velocity gradients in the opposite sense to the CND rotation.
In addition, position velocity cut $b$ shows clumps to the north of
the CND at --15 and +70 km s$^{-1}$ (see also spectrum M in Figure
\ref{hcnspec}) which do not follow the rotation pattern of the CND.
We therefore conclude that the CND is not a disk, but rather a ring
with a definite outer edge in HCN(1-0) and HCO$^+$ emission.

\subsection{Kinematics of CND}

Kinematically, the central $2'$ is largely consistent with a rotating
ring of molecular gas with a peak at a radius of $\sim45''$.  The signature
of the rotation is clearly seen in the spectral peaks at 75 and 100
km s$^{-1}$ at positions B and C to the northeast, and at --97 km s$^{-1}$ at
position I to the southwest. Coherent rotation cannot account for all
of the spectral features seen in the CND (see spectra E, G and H).  To
better address the question of coherent structure and rotation in the
CND, we present in Figure \ref{psvlcnd} a position-velocity diagram of
emission around an ellipse with a major axis of 40$''$, an inclination
of 60$^{\circ}$, and a position angle of 25$^{\circ}$ showing emission
from the CND only. The CND is not a complete ring, with a noticeable
lack of emission in the southeast, corresponding to $\Theta
\sim60^{\circ}$ in the velocity range +60 to --10 km s$^{-1}$.  The CND
appears to consist of either two or three distinct streamers rotating
around the nucleus.  The southwest lobe of the CND is seen in the top
middle of the position-velocity diagram, between $\Theta$=50$^{\circ}$
to 250$^{\circ}$ and velocity range --10 to --130 km s$^{-1}$.  The northeast
lobe of the CND is seen in the lower half of the image from
$\Theta$=0$^{\circ}$ to 60$^{\circ}$ and $\Theta$=325$^{\circ}$ to
360$^{\circ}$ in the velocity range +50 to +170 km s$^{-1}$.  There is a
third, possibly distinct feature, from $\Theta$=160$^{\circ}$ to
325$^{\circ}$ in the velocity range +0 to +125 km s$^{-1}$.  This feature may
or may not be a continuation of the southwest lobe.  The small
northern gap at $\Theta \sim220^{\circ}$ and 0 km s$^{-1}$ is discussed in the
next section.

Previous suggestions have been made that the CND is a warped ring with
inclination angles 70$^{\circ}$ and 50$^{\circ}$ for the southwest and
northeast lobes respectively. The warped structure and the high
velocity dispersion lead to a short dynamical lifetimes of 10$^4$ --
10$^5$ years \citep{gus87}.  \citet{jac93}, propose that the CND is
composed of several dynamically distinct streamers rotating around the
nucleus.  In this case, the two bright lobes are at different
inclinations because they are are on separate orbits.  The
morphological incompleteness of the ring as seen in our HCN(1-0) and
HCO$^+$ data, along with the evidence for separate streamers seen in our
position-velocity diagram of gas around the CND, support the thesis of
\citet{jac93}.

\subsection{Infall from the North}

While there is widespread HCN and HCO$^+$ emission throughout the central
$200''$, there is also a conspicuous lack of emission in the northwest
quarter of the image (at positive galactic latitude), where little dust
or gas has been seen before.  It seems that any prior material in this
region has largely been swept clear, either from an outflow
possibly related to the nucleus, or through accretion towards the
center.

\begin{figure}
\centering
\includegraphics[angle=270,width=6in]{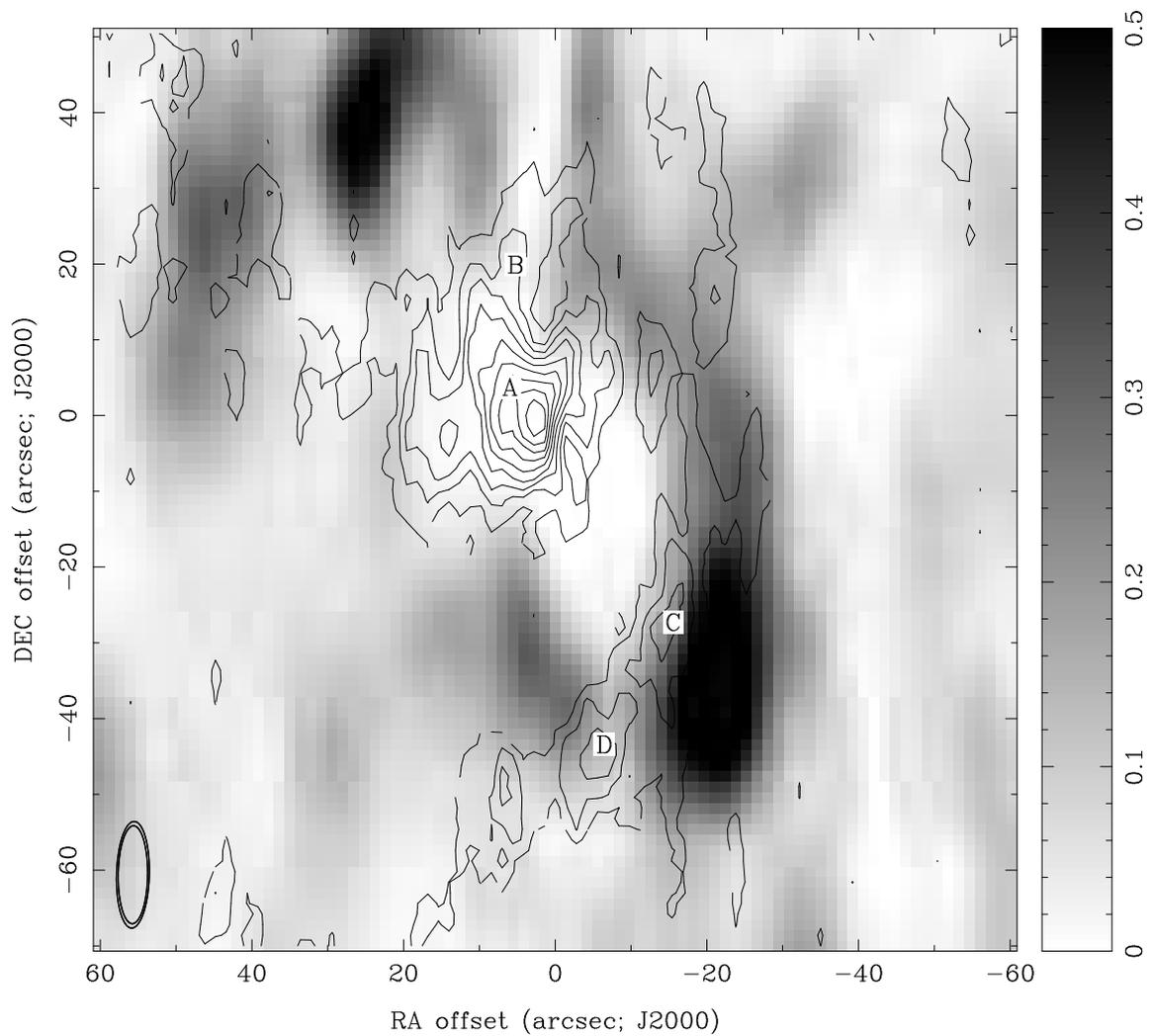}
\caption{H 42$\alpha$ emission integrated from -150 to +150 km
s$^{-1}$.  Contours are at intervals of 12.5 mJy from 12.5 to 125 mJy
beam$^{-1}$.  The grey scale image is the HCN(1-0) emission integrated
from -150 to +150 km s$^{-1}$.  The positions of H 42$\alpha$ spectra
are indicated.  The synthesised beam FWHM is shown in the bottom left
corner. The position of OH maser B from \citet{yus99} is marked by an
``X''.}  \label{h42}
\end{figure}

\begin{figure}
\centering
\includegraphics[angle=270,width=6in]{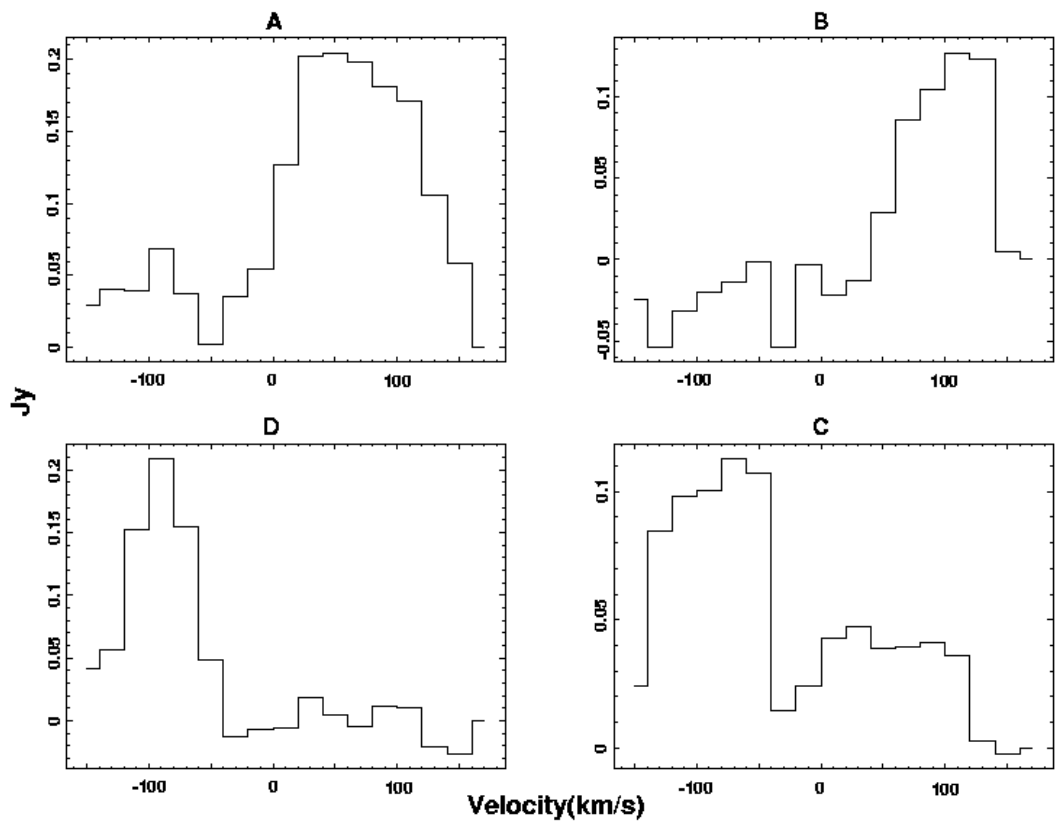}
\caption{H 42$\alpha$ spectra at positions marked in Figure
\ref{h42}.}  \label{h42spec}
\end{figure}

\begin{figure}
\centering
\includegraphics[angle=270,width=6in]{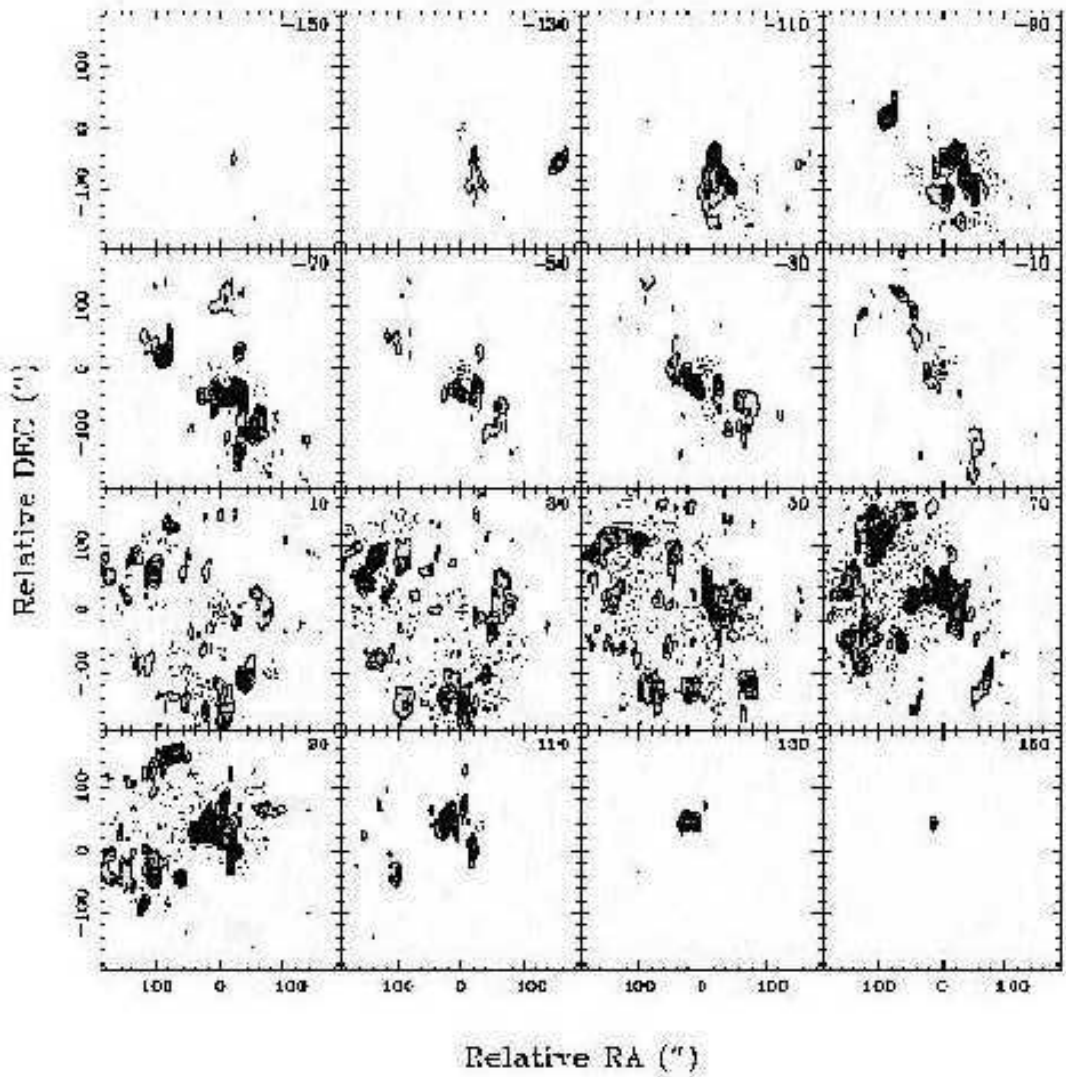}
\caption{Images of HCN(1-0) emission in 20 km s$^{-1}$ intervals.
The contour interval is 0.25 Jy beam$^{-1}$ (0.6 K).}  \label{hcn20}
\end{figure}

\begin{figure}
\centering
\includegraphics[angle=270,width=6in]{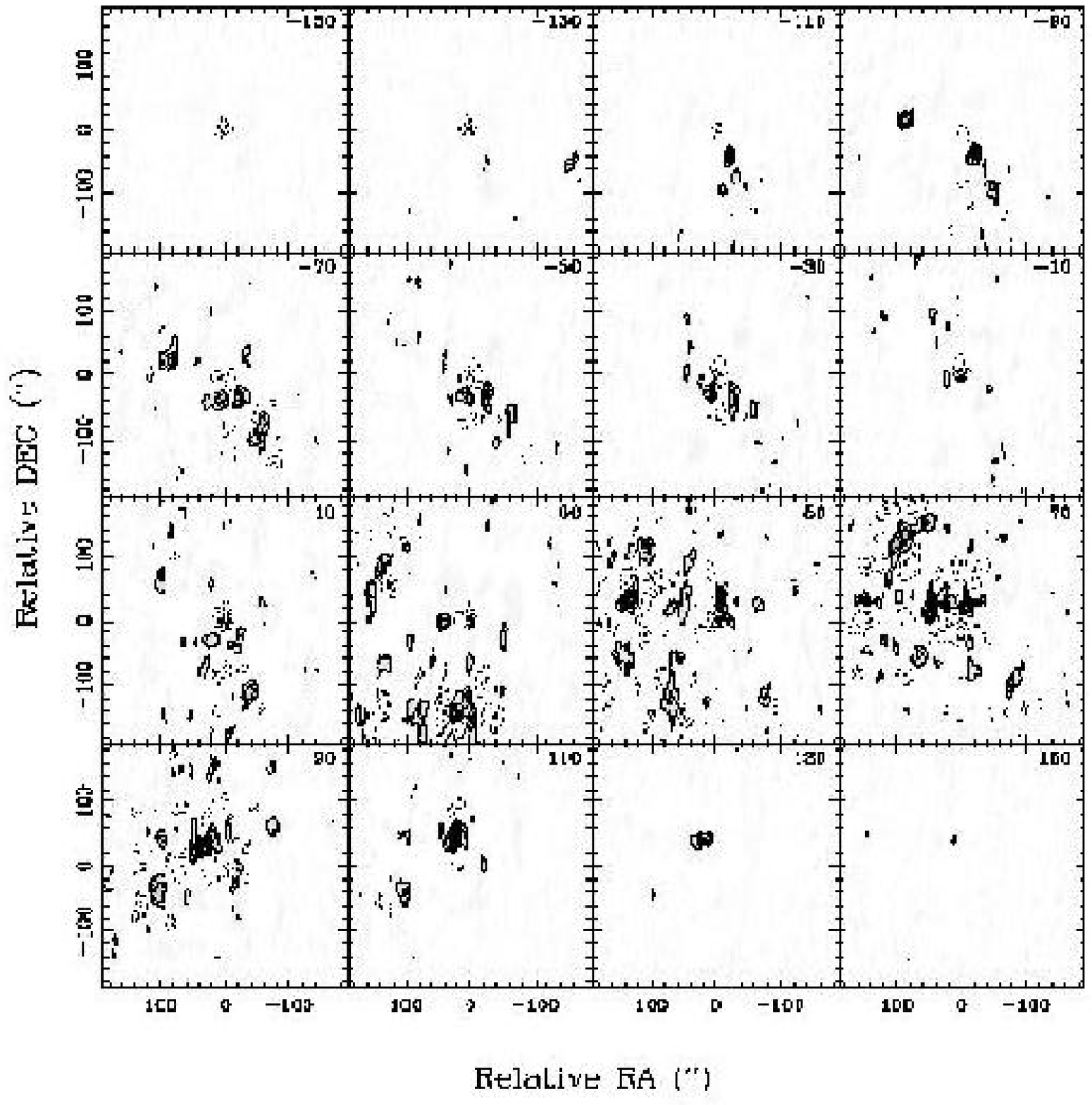}
\caption{Images of HCO$^+$ emission in 20 km s$^{-1}$ intervals.  The
contour interval is  0.25 Jy beam$^{-1}$ (0.6 K).}  \label{hco20}
\end{figure}

\begin{figure}
\centering
\includegraphics[angle=270,width=6in]{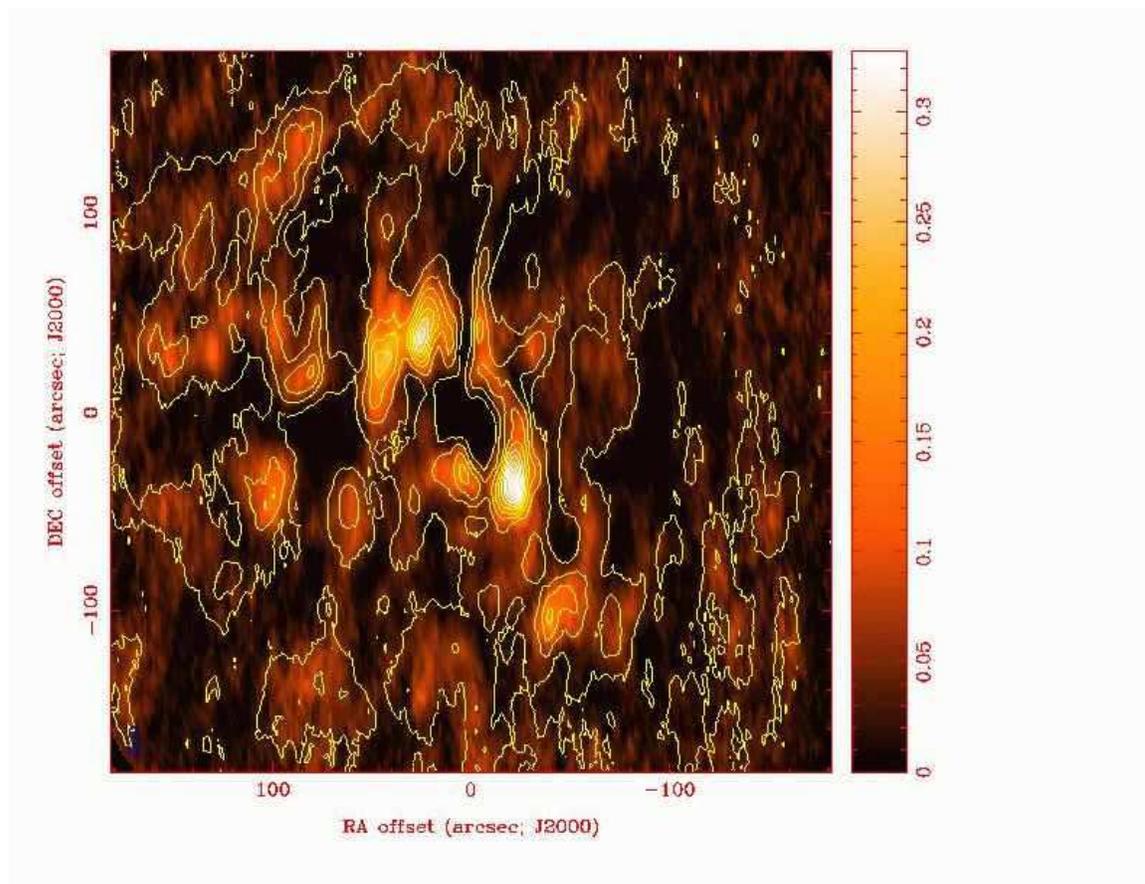}
\caption{Comparison of HCN(1-0) (contours) and HCO$^+$ (grey/color
image) emission. Contours from 0.04 to 0.78 Jy beam$^{-1}$.  (3.5 K/Jy)}
\label{hcn-hco}
\end{figure}

\begin{figure}
\centering
\includegraphics[angle=0,width=6in]{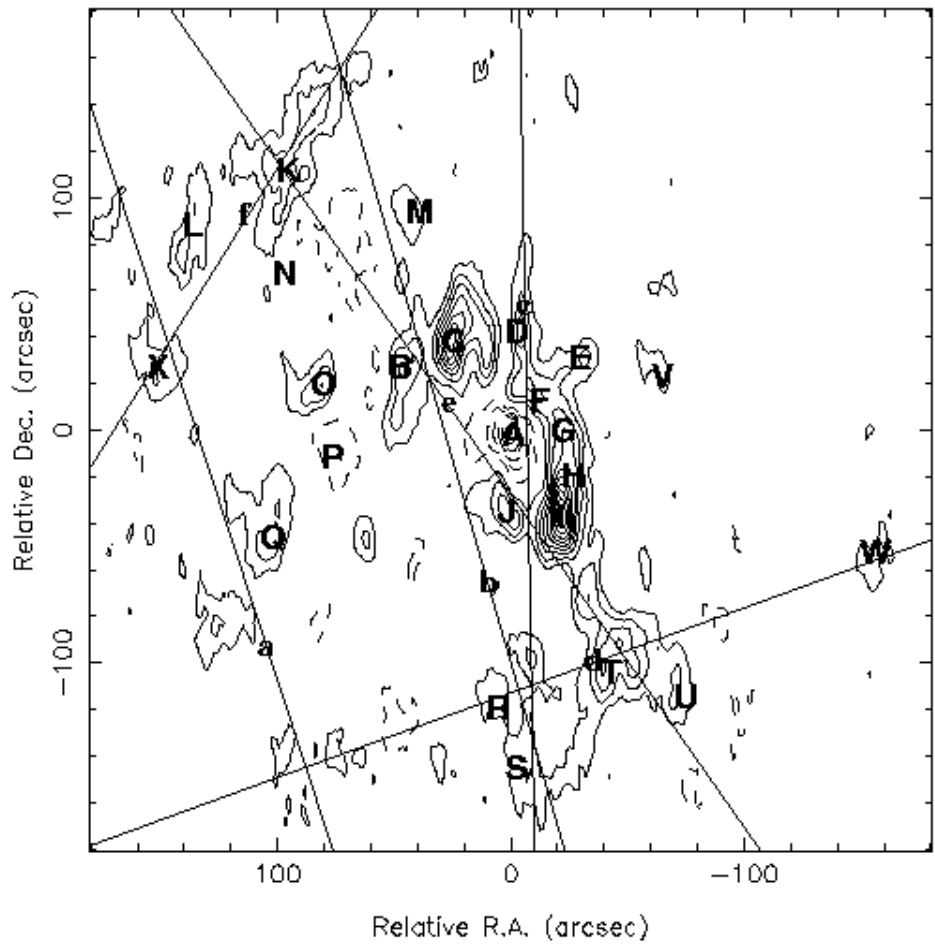}
\caption{HCN(1-0) and HCO$^+$ spectra positions (labeled A to X) and
position-velocity cuts along lines (labeled a to f), superposed on
contours of HCN emission integrated from -150 to 150 km s$^{-1}$. The
contour interval is 0.213 K. } \label{specpos}
\end{figure}

\begin{figure}
\centering
\includegraphics[angle=270,width=6in]{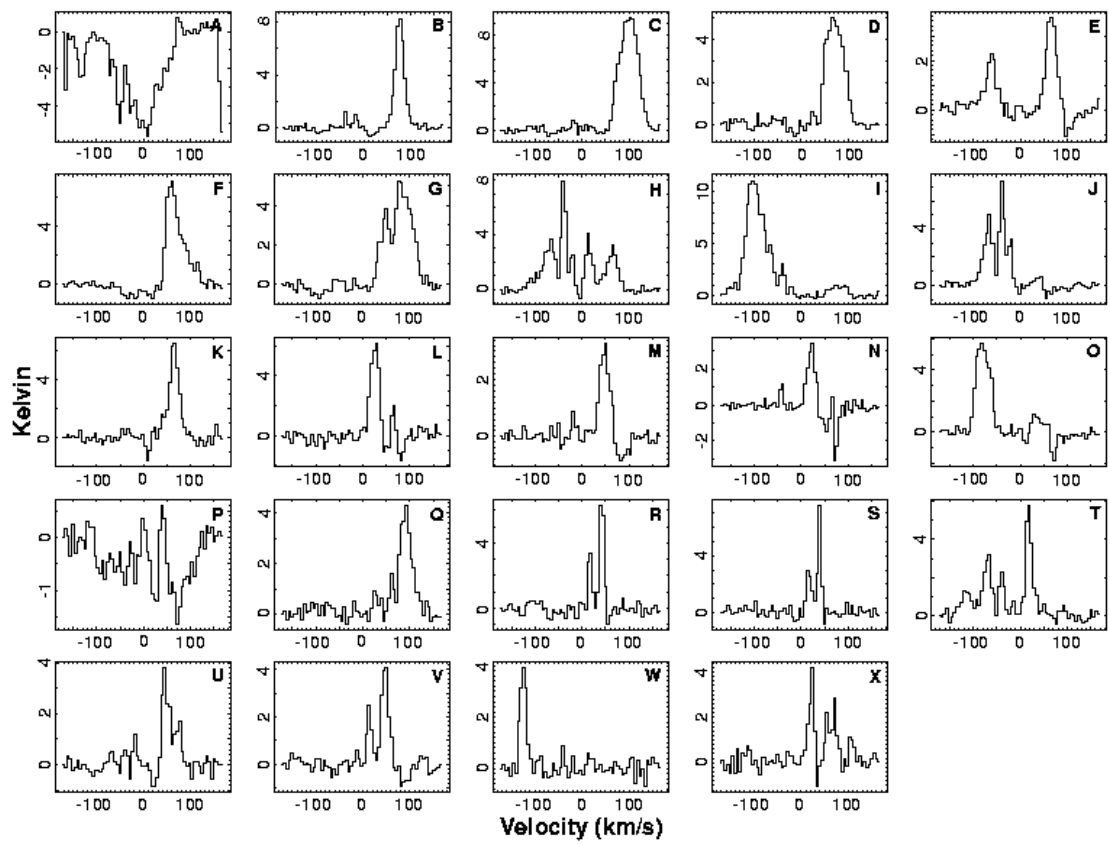}
\caption{HCN(1-0) spectra at positions shown Figure
\ref{specpos}.}  \label{hcnspec}
\end{figure}

\begin{figure}
\centering
\includegraphics[angle=270,width=6in]{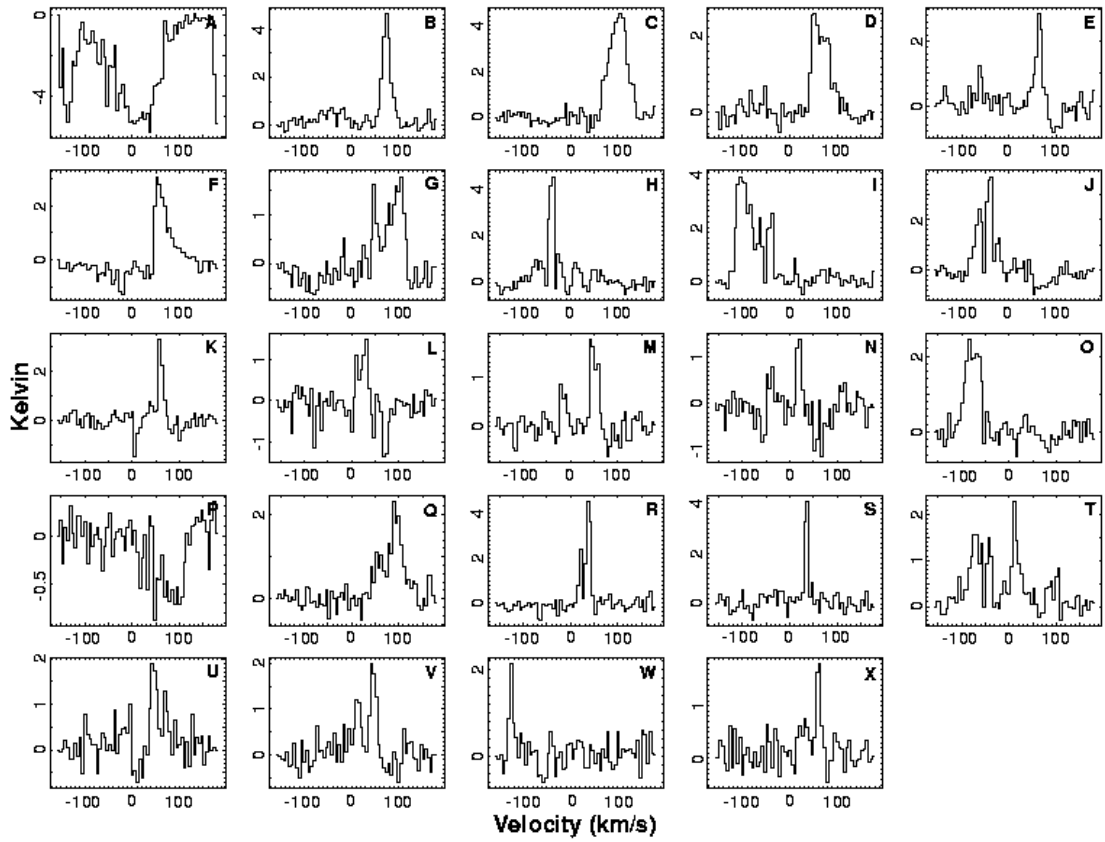}
\caption{HCO$^+$ spectra at positions shown Figure \ref{specpos}.}
\label{hcospec}
\end{figure}

\begin{figure}
\centering
\includegraphics[angle=270,width=6in]{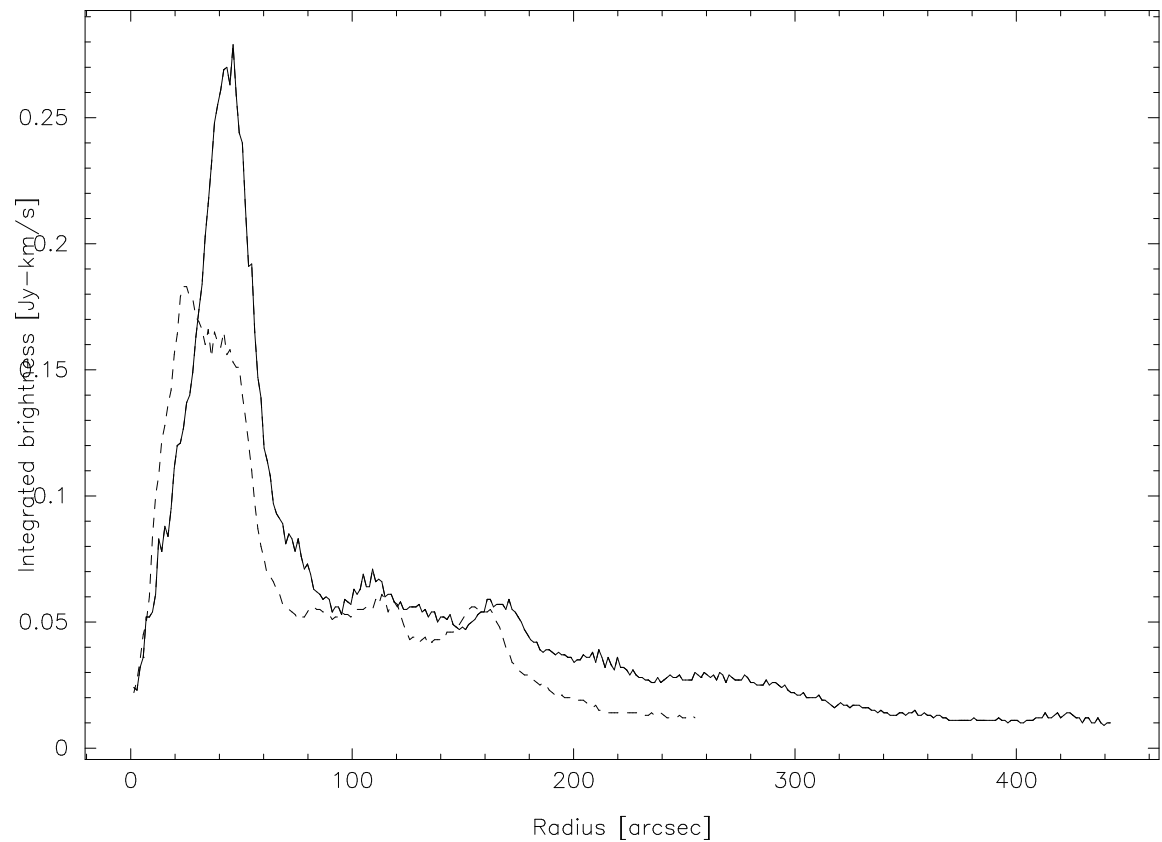}
\caption{Radial profile of HCN(1-0) emission averaged around
elliptical annuli. As it is not clear whether the more extended
emission is in the same plane as the CND (which itself may not lie in
one plane), figure \ref{radprofile} shows radial profiles in the plane
of the sky (dashed line), and de-projected from an inclination of 56$^{\circ}$ (solid line).}  \label{radprofile}
\end{figure}

\begin{figure}
\centering
\includegraphics[angle=0,width=6in]{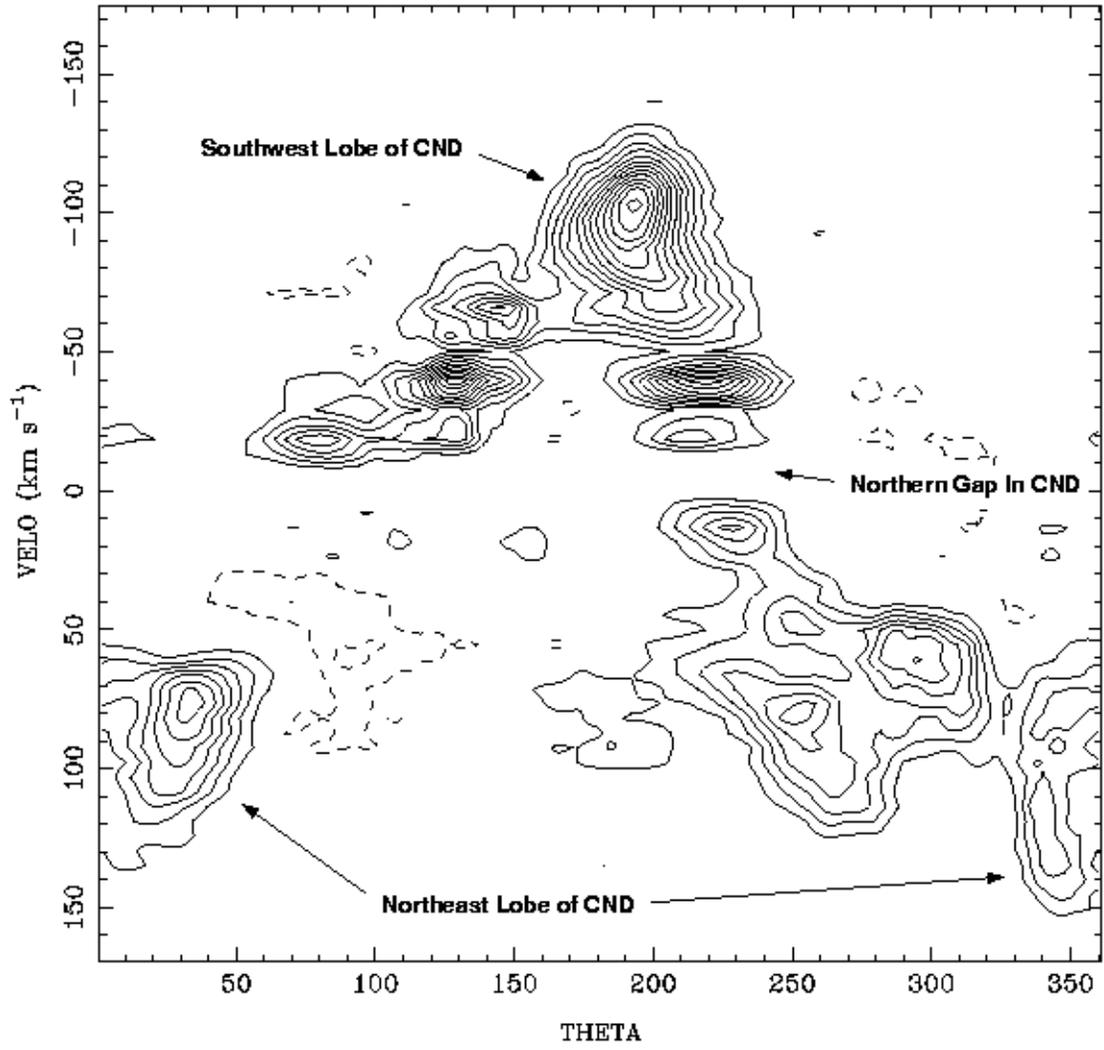}
\caption{Position-velocity diagram of emission in the CND.}
\label{psvlcnd} 
\end{figure}

Figure \ref{h42} shows a distinct, narrow gap in HCN(1-0) and HCO$^+$
velocity-integrated emission in the northern part of the CND exactly
along the northern arm of the mini-spiral.  The HCN and HCO$^+$ spectra
to the east of this gap (spectrum C in Figures \ref{hcnspec},
\ref{hcospec}) peak at 100 km s$^{-1}$ with a FWHM of 45 km s$^{-1}$
which match those in the ionized gas \citep[Figure 11c]{rob93}. To the
west of the gap (spectrum D), the spectra peak at $\sim$70 km s$^{-1}$
with a comparable FWHM.  These linewidths are roughly twice that of
spectra on either side in the CND (spectra B and F).  The increase in
the linewidth of the gas, along with the lack of negative contours,
indicates that the gap is not due to absorption by gas in front of the
CND.  Therefore, the gap seen in the velocity integrated map is a real
feature of the CND.

A 1720 MHz OH maser from collisionally excited gas behind a C-type
shock \citep{yus99} is located in the gap (position X in Figure
\ref{h42}).  The increase in linewidth of the gas near the gap in
addition to the complete lack of HCN(1-0) and HCO$^+$ emission in the gap
and the presence of the OH maser indicate that material is interacting
with the CND at this location.  The superposition of the northern arm
with the gap (see Figure \ref{h42}) suggests that the northern arm
crosses the CND at this point.

It appears that the inflow of material began outside the CND and
disrupted the ring as it fell inward rather than originating from
cloud-cloud collisions in the CND itself. The HCN and HCO$^+$ emission
west of the gap extends north to more than $100''$ radius (Figure
\ref{hcn-hco}) beyond the CND.  Figure \ref{pv} shows position-
velocity diagrams taken from cuts marked in Figure \ref{specpos}.  Cut
$c$ shows that the spur to the north continues at the velocity of the
northern arm of ionized gas. In a model of the far-IR structure at 31.5
and 37.7 micron \citep{lat99}, the northern arm is on a parabolic
orbit 10$^\circ$ from the plane of the CND and extends 1.4 pc to the
northwest of the CND, suggesting that the northern arm is a separate
stream of gas and dust which intersects the CND as it falls in towards
the center.

\subsection{Interaction of Molecular clouds with Sgr A East}

The HCN and HCO$^+$ images in Figures \ref{hcn20} and \ref{hco20} show
several structures outside the CND.  From -90 to +90 km s$^{-1}$ there
is emission throughout the eastern region of the images which may be
associated with the 50 km s$^{-1}$ cloud. There are absorption dips at
both 50 and 70 km s$^{-1}$ in the spectra to the northeast, and also to the
southwest in spectra R, S, and T. Evidently the 50 and 70 km s$^{-1}$ clouds
extend over much of the Galactic center, as also indicated by large
scale imaging of CS J=1-0 emission \citep{tsuboi99}. CS J=1-0 is a
good tracer of dense molecular gas (n$\sim10^4$ cm$^{-3}$). Tsuboi's
Figures 3 and 4 show that the CS emission at 50 to 70 km s$^{-1}$
extends over most of the Galactic center.  In the HCN and HCO$^+$ images,
there is extensive evidence of interaction of Sgr A East with the 50 km
s$^{-1}$ cloud to the east; to the west, outside of the CND, there is
little high-velocity HCN and HCO$^+$ emission, except for an isolated
emission feature at -130 km s$^{-1}$ located $156''$ west, $52''$
south (spectrum W), and HCN emission at -60 km s$^{-1}$ located $30''$
west, $30''$ north (spectrum E). Blue-shifted C+ emission at the
latter location indicates a substantial UV flux density and proximity
to the Galactic center, and corresponds to the northwest edge of Sgr A
East \citep{lug86}.

\begin{figure}
\centering
\includegraphics[angle=270,width=6in]{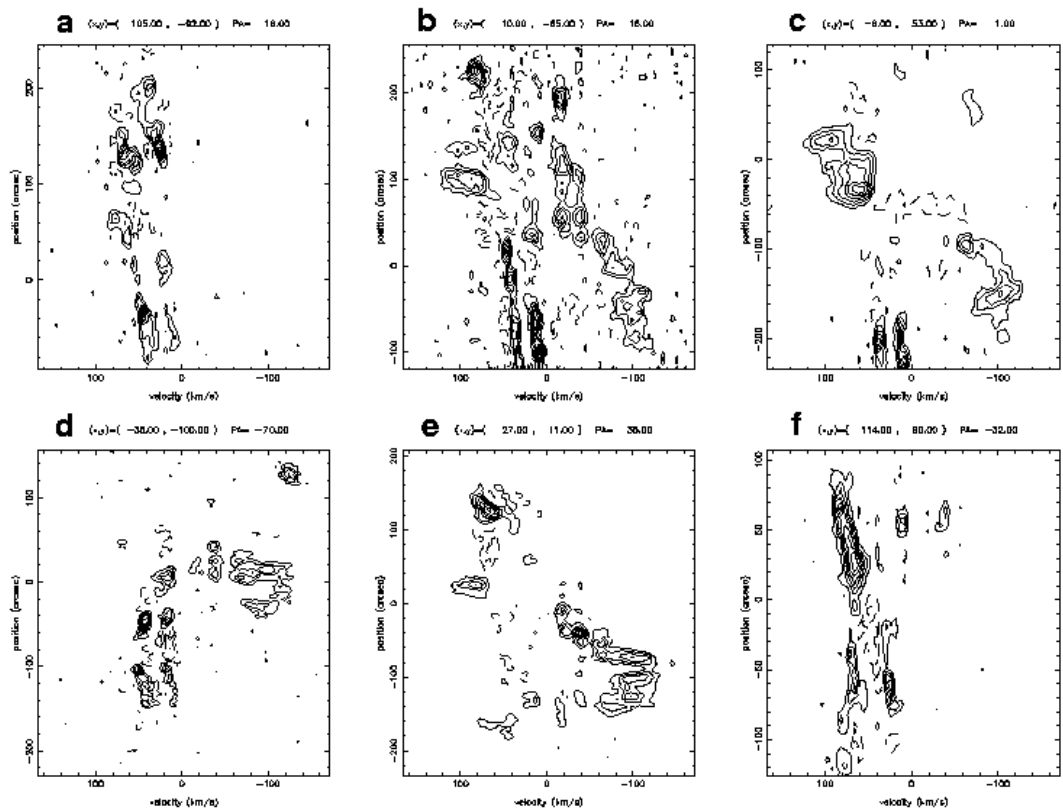}
\caption{Position velocity cuts as indicated in Figure
\ref{specpos}. Positions are measured relative to the location of the
label and north is up in all of the panels.  Cuts $a$ and $b$ run along
the ridges of the NH$_3$ (1,1) emission.  Cut $c$ goes through emission
to the west of the northern gap.  Cuts $d$ and $e$ probe the
high-velocity emission to the southwest of the CND. Cut $f$ runs though
the 50 km sec$^{-1}$ cloud which borders Sgr A East in the north east.}
\label{pv}
\end{figure}

A number of studies show that Sgr A East is expanding into the 50 km s$^{-1}$
cloud \citep{mez89,zyl90,gen90,hoe91,ser92,coi00}.
\citet{ser92} mapped a dense molecular ridge in
CS J=7-6 and 5-4 emission, which wraps around the eastern side of Sgr A
East.  Figure \ref{hcn-nh3-oh-6cm} shows that NH$_3$ emission also
closely follows the eastern edge of Sgr A East \citep{coi00}.  High velocity red- and
blue-shifted emission is seen along the ridge in CS, HCN(1-0) and
HCO$^+$.  Figures \ref{hcn20} \& \ref{hco20} show blue-shifted emission at
--50 to --90 km s$^{-1}$ located $\sim90''$ east and $\sim20 - 40''$ north, and
at --50 to --30 km s$^{-1}$ in the arc of emission in northern part of the
images at 90$''$ east and 140$''$ north.  Red-shifted emission is seen
at +90 to +110 km s$^{-1}$ along the molecular ridge, with the highest
velocities at 100$''$ east and 40$''$ south.  In Figure
\ref{hcn-nh3-oh-6cm} the high-velocity emission, comprised of several
clumps of gas, generally lies inside the molecular ridge of compressed
gas imaged in CS J=7-6 and J=5-4 \citep{ser92}.  Serabyn et al. argue that
the high-velocity gas has been accelerated by the impact of Sgr A East
with the 50 km s$^{-1}$ cloud. Since both red- and blue-shifted gas is
observed, parts of the 50 km s$^{-1}$ cloud must lie on both the front and
back side of the Sgr A East expanding shell.  Spectra K, L, N, O, and Q
show the high-velocity emission from this region. Clumps of HCN and
HCO$^+$ emission, with steep velocity gradients (e.g. spectra O and Q at
--80 and +100 km s$^{-1}$ respectively), are superposed on the smoother,
highly resolved ambient cloud material at 20 to 60 km s$^{-1}$ with more
modest velocity gradients of ~5-8 km s$^{-1}$ arcmin$^{-1}$ (see position-
velocity profile $b$ in Figure \ref{pv}). These data suggest that clumps
of gas have been compressed and accelerated as Sgr A East has expanded
into the ambient cloud material.  The ambient cloud material is self-
absorbed in HCN and HCO$^+$ emission, indicating warmer gas 
overlaid by colder gas.

Figure \ref{hcn-nh3-oh-6cm} marks the positions of OH (1720) masers
detected close to the Galactic Center superposed on an image of 6 cm
continuum emission which shows the Sgr A East shell.  These 1720 MHz
masers, which are seen without the main transitions, are usually
associated with the interaction of expanding supernova shells with
molecular clouds \citep{yus99}. The velocity and velocity gradients
seen in the OH masers, $\sim$2 km s$^{-1}$ pc$^{-1}$, are similar to
those observed in the 50 km s$^{-1}$ cloud. \citet{yus99} argue that
the OH masers arise from the interaction of Sgr A East with the 50 km
s$^{-1}$ cloud which is located within 5 pc behind the Galactic
center.  Absorption of the continuum radiation from Sgr A East by Sgr A
West at cm wavelengths \citep{yus87, ped89}, locates Sgr A
East behind Sgr A West. The red and blue-shifted gas $100''$ to the
east suggest that the 50 km s$^{-1}$ cloud and Sgr A East are at about 
the same distance.
Position-velocity cut 'f' in Figure \ref{pv}
traces gas in the northern part of the 50 km sec$^{-1}$ cloud, where it wraps
around the expanding shell of Sgr A East.  Emission in the upper
half of the figure is red-shifted to +60 to +80 km sec$^{-1}$ and 
has a narrower linewidth than the gas near the core of the GMC,
seen in the lower half of the diagram from --20 to --80 arcsec.
We do not see corresponding blue-shifted emission in the northern
part, which implies that this gas must be located {\it behind} and
adjacent to Sgr A East.

Figure \ref{hcn-nh3-oh-6cm} shows a progression to the east of  
the Galactic Center of high-velocity HCN(1-0) and HCO$^+$, followed by the ridge
of NH$_3$, and a string of HII regions.  The high-velocity gas is
accelerated and shocked by Sgr A East, which may be comprised of several
expanding supernova shells. The HII regions are
at a larger radius than the the current wave of compressed gas, and
must trace an earlier epoch of star formation.

\subsection{Comparison to NH$_3$}

\begin{figure}
\centering
\includegraphics[angle=270,width=6in]{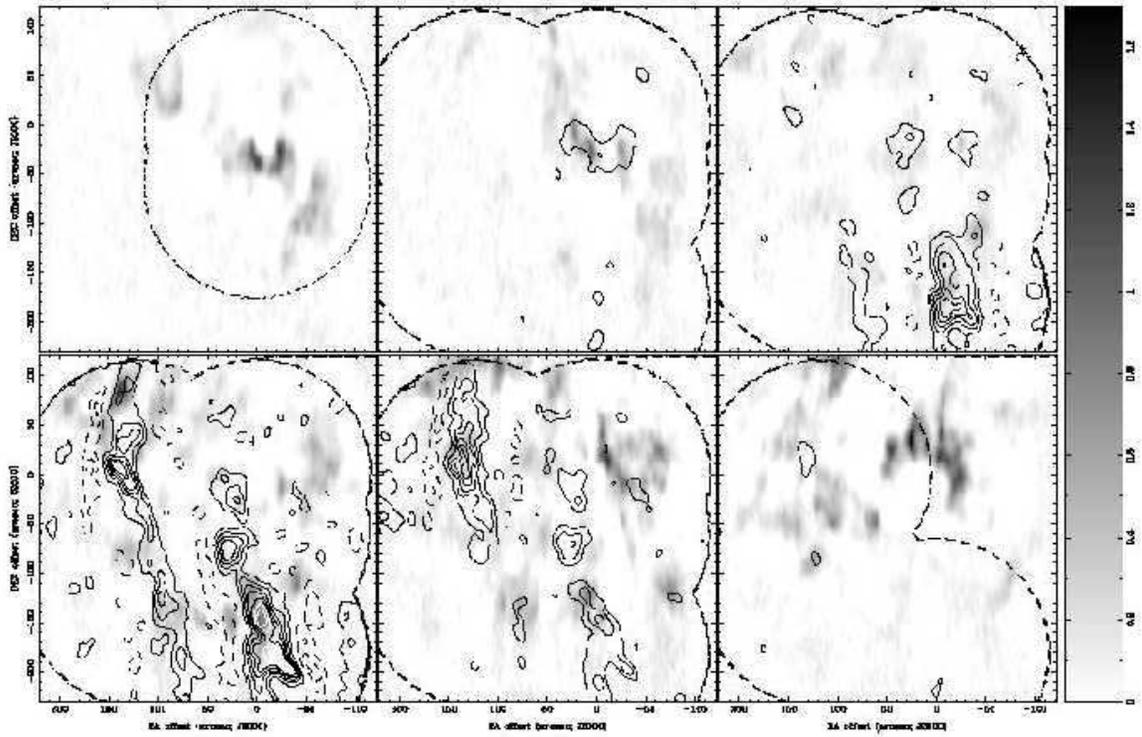}
\caption{Comparison of HCN(1-0) (greyscale) and NH$_3$(1,1)
\citep{coi00} (contours) in 6 velocity intervals:  top left -75 to -35
km s$^{-1}$, Top middle -35 to -10 km s$^{-1}$, top right -10 to 15 km
s$^{-1}$, bottom left 15 to 40 km s$^{-1}$, bottom middle 40 to 600 km
s$^{-1}$, and bottom right 60 to 100 km s$^{-1}$.}  \label{hcn-nh3}
\end{figure}

The NH$_3$(1,1) and (2,2) distributions are very similar
\citep{coi99,coi00}, as are the HCO$^+$ and HCN(1-0). The comparable
angular resolution and spatial coverage of these data sets makes a
comparison of NH$_3$ to HCN and HCO$^+$ results of immediate
interest.  However, the velocity range and spatial coverage of
the NH$_3$(1,1) data do not match that of the HCN(1-0) data.  HCN(1-0)
and HCO$^+$ were observed over the velocity range --150 to +150 km
s$^{-1}$ in a mosaic pattern covering the central $5'$.  The VLA
observations of NH$_3$(1,1) and (2,2) were obtained in 5 fields.  The
central and northern field cover --75 to +55 km s$^{-1}$; the southern
and two eastern fields cover --40 to +90 km s$^{-1}$.  Figure
\ref{hcn-nh3} plots HCN(1-0) and NH$_3$(1,1) emission in six velocity
ranges covering the total velocity interval of --75 km s$^{-1}$
to +90 km s$^{-1}$.  The single outer contour shows the spatial extent
of the mosaic primary beam pattern for the NH$_3$(1,1) data in each 
velocity range.

\begin{figure}
\centering
\includegraphics[angle=0,width=6in]{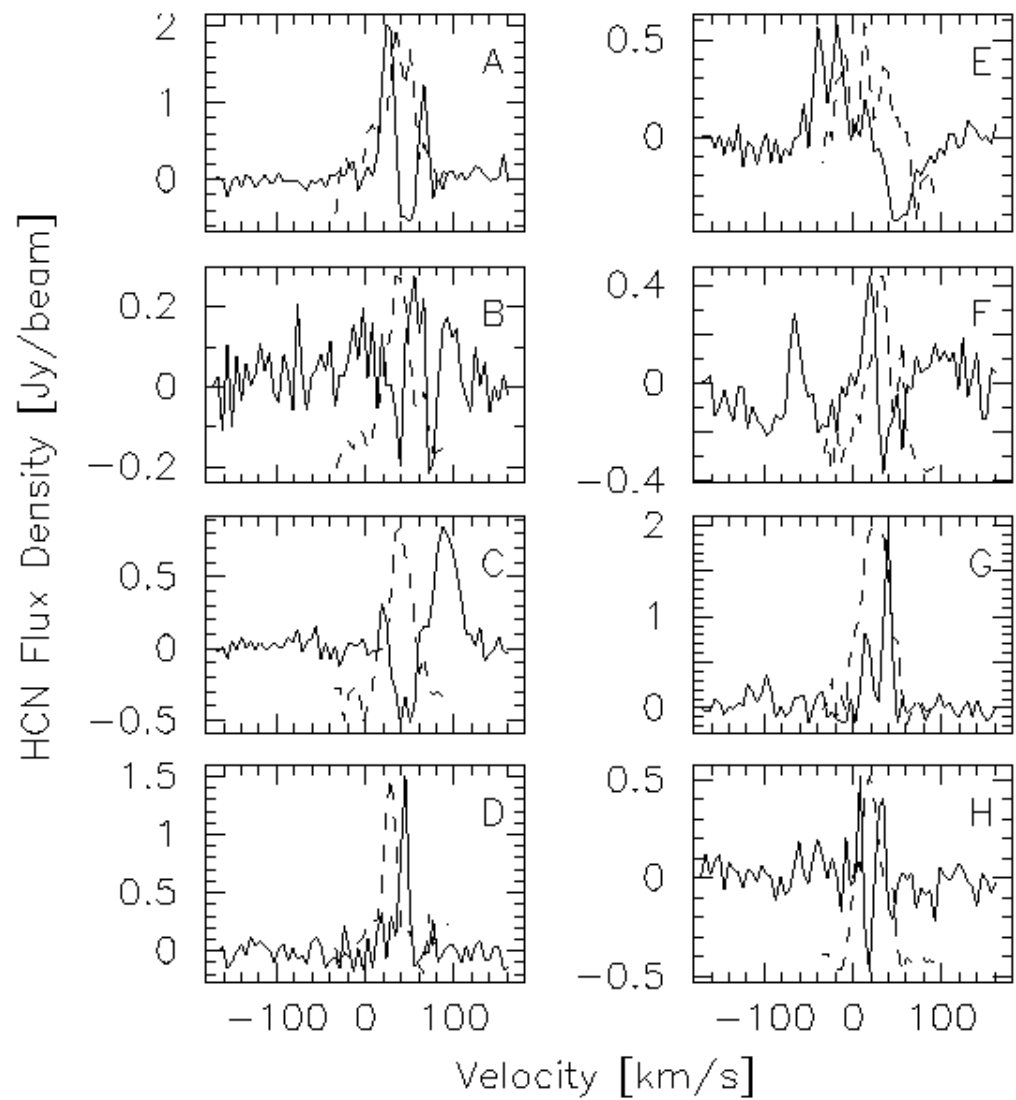}
\caption{Spectra of HCN (solid lines) and NH$_3$(1,1) (dashed
lines) at the positions marked in Figure \ref{hcn-nh3}.}
\label{hcn-nh3spec}
\end{figure}

\begin{figure}
\centering
\includegraphics[angle=270,width=6in]{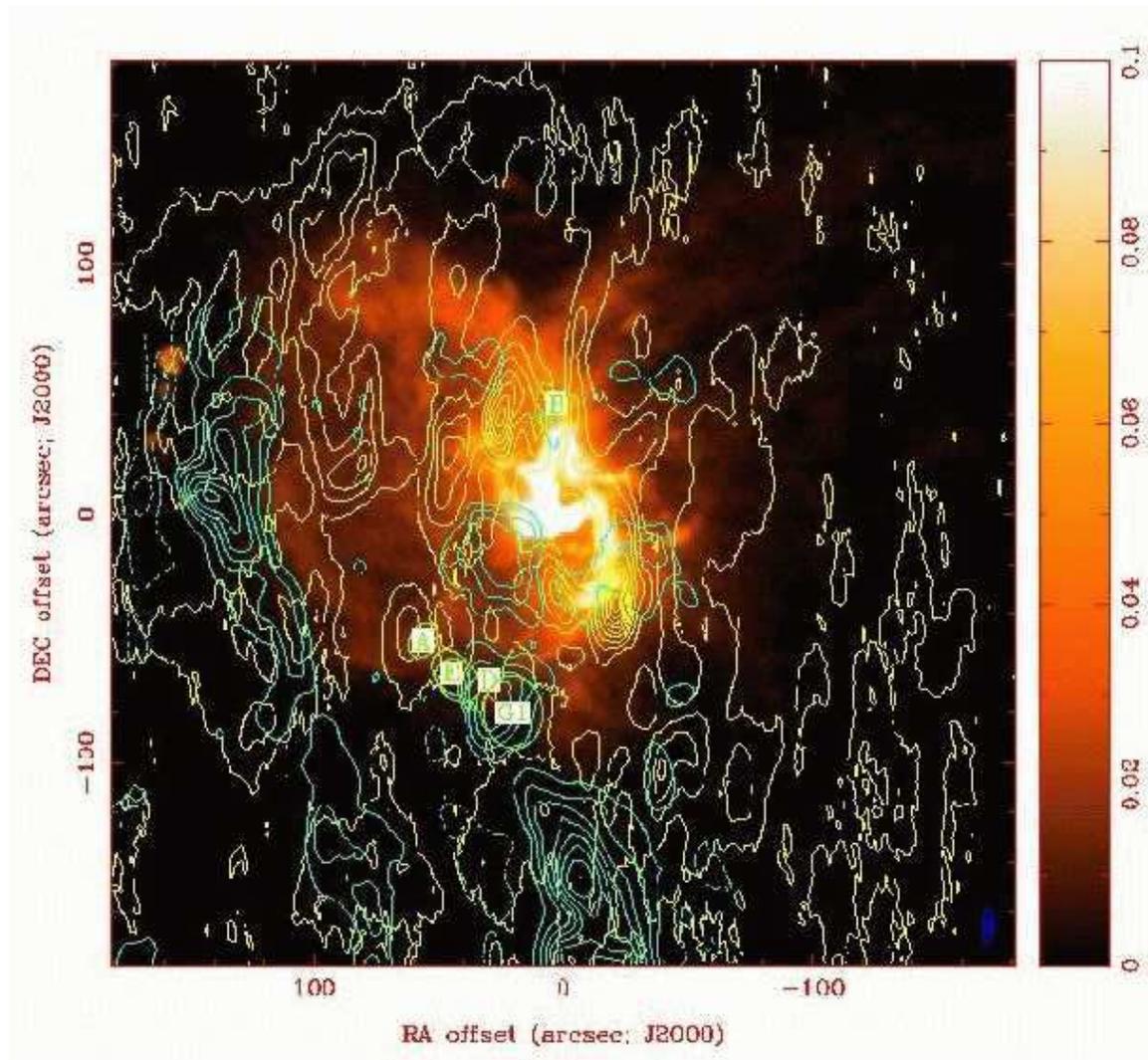}
\caption{Comparison of HCN(1-0) (yellow contours) and NH$_3$(1,1)
\citep{coi00} (blue contours) in the velocity range --75 to +55 km
s$^{-1}$ overlaid on 6 cm continuum emission (color).  Positions of OH
masers \citep{yus99} are marked with letters.}  \label{hcn-nh3-oh-6cm}
\end{figure}

The NH$_3$ emission is brightest in two parallel ridges running from
the northeast to southwest across the mosaic in the velocity range of
0 to 60 km s$^{-1}$ as seen in \citet{coi00}.  Figure \ref{hcn-nh3}
shows patchy HCN(1-0) emission in this region, but these ridges are
not as apparent in HCN as in NH$_3$.  Another difference between the
HCN and NH$_3$ emission is that the NH$_3$ does not trace the high-
velocity features seen in the CND in HCN(1-0).  This is most noticable
in the top left and bottom right panels where the HCN shows strong
emission from the CND while there is no indication of NH$_3$ emission
at these velocities.  Figure \ref{hcn-nh3spec} compares the spectra of
HCN(1-0) and NH$_3$(1,1) at the positions of the NH$_3$(1,1) peaks
from \citet{coi00}.  The NH$_3$(1,1) spectra have been scaled and
shifted to match the height of the HCN line and are shown for
comparison of the line shape only and not the absolute flux level.  In
most of these spectra, a deep absorption feature is seen in HCN(1-0)
at the position of the peak NH$_3$(1,1) emission.

{\it Effects in the GMCs}  

The high abundance and high optical depths of HCN(1-0) and HCO$^+$(1-0)
make these tracers particularly susceptible to absorption along the
line of sight as compared to NH$_3$.  Gas along the line of sight that
is not related to the Galactic center and is at a lower column density
and temperature may cause absorption. This can occur either because of
a lower physical temperature or a lower volume density leading to a
lower excitation temperature.  In examining the spectra in Figure
\ref{hcnspec}, it is clear that most of the narrow absorption dips are
due to intervening matter which are not located at the Galactic
Center.  These dips are not seen in the higher transitions of HCN
\citep{mar95} which suggests that the absorbing material is probably
cold and at a low enough density that the upper levels are not
populated.  The absence of the dips in NH$_3$ is probably due to its
relatively lower abundance and hence lower optical depths.  However,
even in NH$_3$, absorption can be seen in deep integrations
\citep{ser86}.

The high optical depth of the HCN(1-0) and HCO$^+$ lines can also lead to
self-absorption by cool outer layers of a cloud.  This self-absorption
may account for the lack of HCN(1-0) and HCO$^+$ emission from the more
massive GMCs.  Position-velocity cuts $a$ and $b$ (Figure
\ref{specpos}) run along the ridges seen in NH$_3$(1,1) and (2,2)
emission.  The double-peaked profiles seen in Figure \ref{pv} from 0
to 60 km s$^{-1}$ indicate that the HCN(1-0) emission is self-absorbed
along the ridges of NH$_3$ emission.  Self-absorption to the south of
the CND can also be seen in position-velocity cuts $c$ and $d$.  Cut
$f$ shows additional self-absorption to the northeast in the region of
the 50 km s$^{-1}$ cloud core (but not in the northern part of the ridge,
where the gas is red-shifted).
The location of the NH$_3$ emission in the absorption dips of the
HCN(1-0) emission (Figure \ref{hcn-nh3spec}) is a signature of
self-absorption in the HCN(1-0).  The HCN and HCO$^+$ spectra are highly
correlated with each other, and the HCO is also highly self-absorbed.

From the strength of the dips due to line of sight absorption
and self-absorption we estimate that absorption by intervening
material may lead to underestimates of HCN emission by a factor of
$\sim$2, while self-absorption by the outer parts of clouds can have a
stronger effect of up to a factor of $\sim$5 underestimation of the emission.

{\it Effects in the CND}

Position velocity cut $b$ also shows emission from the CND from --100
km s$^{-1}$ at a position of --20$''$ to +120 km s$^{-1}$ at a position
of +100$''$.  This emission does not appear to be self-absorbed which
implies that the optical depth is lower in the CND than in the GMCs.
The high optical depth of HCN(1-0) in the Galactic Center may explain
why it is brightest in the CND.  Gas in the CND typically has very
broad ($\ge$50 km s$^{-1}$) line widths.  For a very optically thick
line, the line wings will also be optically thick and the broadened
line will appear brighter in total flux than an optically thick but
narrow line.  However, for an optically thin tracer, the broadening in
the CND will only make the line weaker as the emission that was
originally confined to a narrow velocity range is spread out into a
broad line.  In that case, the total flux remains the same.  Although
opacity may not account entirely for the differences in HCN(1-0) and
NH$_3$ emission, it is clear that HCN images will preferentially show
line broadened emission from the CND while emission from the GMCs will
be relatively weaker due to both the narrow line width and
self-absorption.

The differences in the NH$_3$ and HCN maps may also be due to a variety
of chemistry and temperature effects in the region.  Using a large
velocity gradient (LVG) model, \citet{mar95} derive an HCN abundance of
$\sim$10$^{-9}$ with a molecular density n[H$_2$] ~ $\sim$10$^6$
cm$^{-3}$ from a comparison of HCN(4-3) and (3-2) emission in the CND,
consistent with normal abundances.  However, the analysis of just two
transitions which may be optically thick is not very sensitive to
abundance enhancement.  From a comparison of HCN J=1-0, H$^{13}$CN
J=1-0, and HCN J=3-2 \citep{jac93}, \citet{mar93} derive an HCN J=1-0
opacity of 2-4 in the CND, with T$_{gas}\sim$250K and
n[H$_2$]$\sim$10$^6$ cm$^{-3}$. The derived abundances of HCN and HCO$^+$
are $\sim$ 8 10$^{-8}$ and $\sim$10$^{-9}$ which are enhanced by at
least an order of magnitude over typical values.  This abundance
enhancement could also account for the relative brightness of the CND
in HCN and HCO$^+$ compared to other tracers.

NH$_3$ does not appear to be greatly enhanced in the southern streamer
where \citet{coi99} found that the brightness of the NH$_3$ was
consistent with a ``typical'' abundance of NH$_3$/H$_2\approx10^{-8}$.
The relative weakness of NH$_3$ emission from the CND may also
indicate that the NH$_3$ is underabundant in the ring in addition to
the the opacity effects discussed above.  The small amount of NH$_3$
emission that does appear to come from the CND is found in
low-ionization regions on the outer edge of the CND, with modest
temperatures of 60--70 K \citep{coi00}.  This result matches the dust
temperature of 76 K at a radius of about $50''$ \citep{lat99}.  These
cooler regions of the CND are shielded from UV photons which could
destroy the NH$_3$ molecule.  While NH$_3$, HCN and HCO$^+$ are all
dissociated by ionizing radiation from the center, HCN and HCO$^+$ may be
more efficiently reformed than NH$_3$ in the post-shock chemistry
as a result of cloud-cloud collisions in the CND. 
This could enable the abundances of HCN and HCO$^+$ to remain high in the
CND.
If the CND is comprised of infalling gas on parabolic orbits, the
infall time from a radius of 1.4 pc at a velocity 140 km sec$^{-1}$ is
$\sim$10$^{4}$ yrs.
Models of the chemical evolution for the Orion Hot Core (e.g.
\citep{cha92}), predict a decreased NH$_3$ abundance and an enhanced HCN abundance
after $\sim$10$^{4}$ yrs in an environment similar to the CND at 1 pc
radius.
The grain mantles may be completely removed by this time and the gas
phase NH$_3$, no longer replenished by mantle evaporation, is
destroyed by UV photons in the inner CND environment.

\subsection{Comparison with 1.2mm Dust}

\begin{figure}
\centering
\includegraphics[angle=270,width=6in]{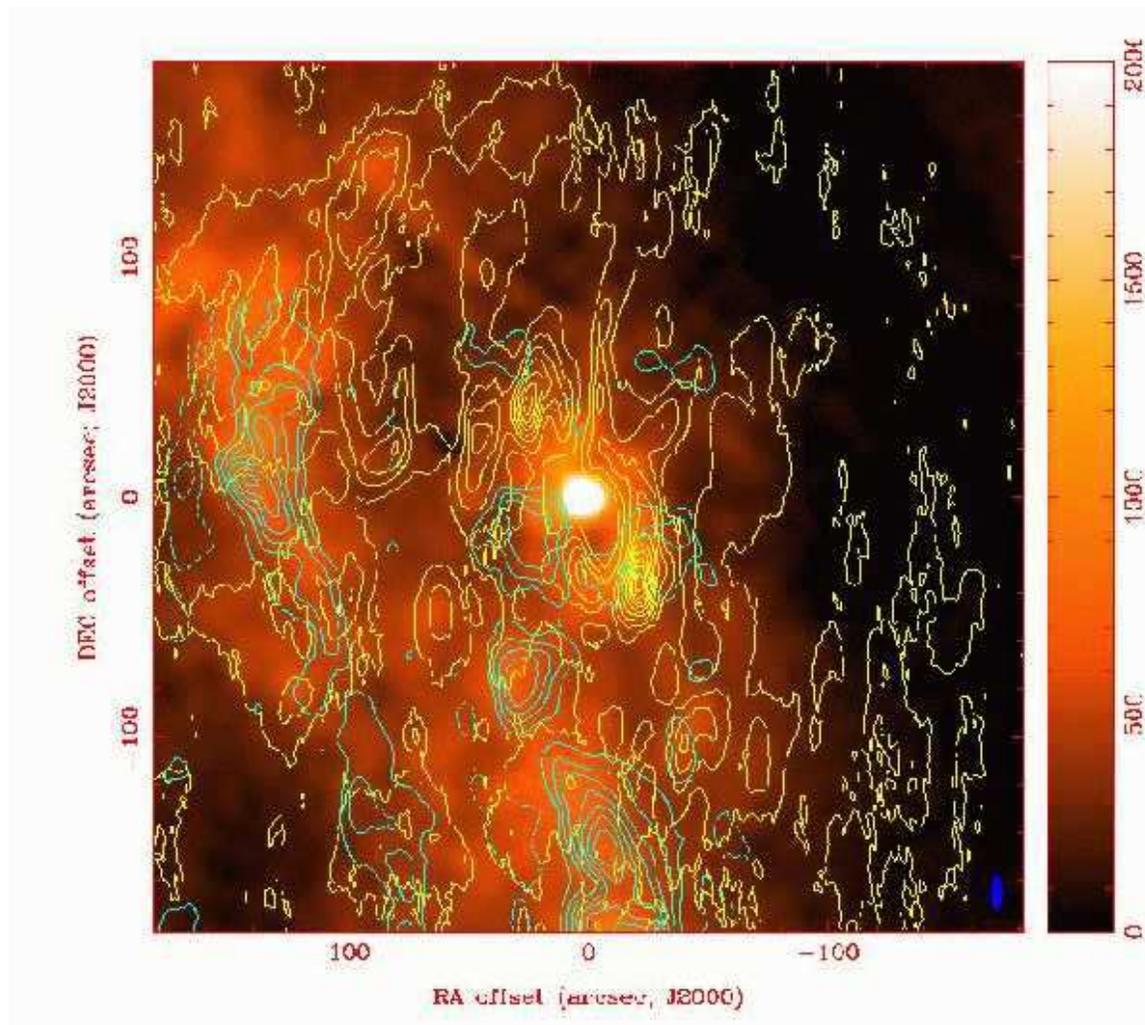}
\caption{Comparison of HCN(1-0), 1.2mm continuum \citep{zyl98},
and NH$_3$(1,1). The 1.2 mm continuum (color background) largely traces
dust. The HCN (yellow contours, levels .04, .125, .21,
.29, .37, .47, .54, .62, .71, .78 Jy), is averaged over a 300 km
s$^{-1}$ interval.  NH$_3$(1,1) emission \citep{coi00} (blue contours),
is averaged over -40 to +55 km s$^{-1}$.}
 \label{hcn-dust}
\end{figure}

Figure \ref{hcn-dust} shows
contours of velocity-integrated HCN and NH$_3$ emission overlaid
on an image of the continuum emission at 1.2 mm \citep{zyl98}.  The
center of the 1.2 mm image is dominated by free-free and dust emission
associated with ionized gas in Sgr A West.  Outside the center, the 1.2
mm image traces emission from dust with column density
3-8$\times10^{23}$ cm$^{-2}$.  The two ridges of NH$_3$ emission agree
well with the dust emission indicating that dust and NH$_3$ are
related. 
The HCN and HCO$^+$ emission agree well with the dust
emission when the effects of self-absorption are considered.

The eastern ridge of NH$_3$ (1,1) emission is coincident with the
eastern edge of Sgr A East where it is impacting the 50 km s$^{-1}$
cloud.  In addition, a supernova shell to the south of Sgr A East was
suggested by \citet{coi00}, centered at $\sim \Delta_{RA}=80'',
\Delta_{DEC}=-120''$.  The shell can be seen as wispy 6cm emission
which seems to impact the southern side of Sgr A East (see Figure
\ref{hcn-nh3-oh-6cm}). The western edge of this shell is coincident
with the ``southern streamer'' seen in NH$_3$ emission.  A string of
OH masers ($A, E, D, G1$) coincides with the bright middle clump of
the ``southern streamer.''  NH$_3$ emission is seen outside these
regions, so shock excitation probably does not account for all of the
observed NH$_3$ emission.  However, NH$_3$ emission in the molecular
ridge and ``southern streamer'' may be enhanced in locations where
expanding shells are pushing into dense molecular gas.

A possible explanation of the correlation between NH$_3$ and dust
emission and the location of shocks is that the NH$_3$ lies on the
edge of GMCs where shocks are heating the grain mantles.  In a
dynamical model of nitrogen molecules in molecular clouds
\citet{nej90} found that NH$_3$ may be formed and stored on grains and
subsequently released in shocks thus providing the gas phase NH$_3$.
In this thin outer layer of the cloud, the grains release NH$_3$ as
they are heated and disrupted, thus enhancing the abundance of NH$_3$.
The NH$_3$ will be warmer than the unperturbed dust in the interior
portions of the cloud, therefore giving a higher excitation
temperature than the dust.  This scenario requires that the
layer of enhanced NH$_3$ abundance is thin so that the total brightness of
NH$_3$ is consistent with an abundance of 10$^{-8}$ averaged
throughout the entire cloud \citep{coi99,coi00}.
It should also be noted that \citet{ter00} found that the
observed abundances of NH$_3$ can be produced using only ion-molecule
chemistry.  Thus the extent to which the dust grains play a role
cannot be determined.

Unfortunately, the present data do not enable us to distinguish
between the theories proposed above.  It is possible that all of these
effects have roles of various importance in the region.

\section{Conclusions}

i) We present images of HCN(1-0) and HCO$^+$ emission and absorption which
trace dense molecular gas in the central 12 parsecs of the Galaxy.  The
HCN(1-0) and HCO$^+$ emission primarily traces filamentary structures
associated with the CND and nearby GMCs; low-level, extended emission
from partially-resolved larger structures is also seen.

ii) The CND appears as a bright, well-defined ring of emission $20'' - 60''$
from the center, at an inclination of 60--70$^{\circ}$.  Although the outer edge of
the ring is not as sharply defined as the inner edge, the CND does not extend
beyond $100''$ in HCN(1-0) and HCO$^+$ emission.  The CND does not appear to be an 
equilibrium structure;  rather, it consists of two or three separate 
streamers in rotation around  the nucleus.  The CND may exist long-term 
in a fluctuating sense, as gas streamers feed the inner parsec.

iii)  Outside the CND, HCN(1-0) and HCO$^+$ trace several filamentary structures
which do not follow the rotation pattern of the $45''$ ring.  High-velocity
emission is seen $90''$ east of Sgr A*, just inside the dense molecular
ridge of shocked material seen at the intersection of Sgr A East
and the 50 km sec$^{-1}$ cloud.  A stream of emission is also seen extending
from the ionized northern arm, through the CND to high-velocity
red- and blue-shifted material to the north. This material may be
falling into the center on an orbit which intersects the CND streamers.

iv) Comparison of the HCN(1-0) and HCO$^+$ with NH$_3$ and dust emission
shows that NH$_3$ emission is well correlated with self-absorbed HCN(1-0) and HCO$^+$
emission, and dust ridges in the nearby GMCs.

v) HCN(1-0) and HCO$^+$ emission appear to be enhanced in the CND relative to
NH$_3$. Further observations of other spectral lines are required to distinguish
between opacity, excitation and chemical explanations for these differences.

These observations show that the accretion of material into the Galactic
Center region may be a chaotic and transient phenomenon.  From the spatial
distribution of the nearby gas with respect to Sgr A East and other nearby
supernova remnants, and from the overall kinematics and especially the
isolated pockets of high-velocity gas, we have the impression that the
nearby molecular clouds are being disrupted by the expanding shells
associated with Sgr A East.  Infall of material appears to approach the
CND from a number of different directions on different orbits.  Sgr A East
itself may be driven by multiple outflow events associated with the
central black hole or multiple supernovae.  We favor a model where the
accretion process is highly non-steady and non-uniform at least on the
timescale of the rotation period of the CND, 10$^{5}$ yrs.  In this
scenario, the CND does not appear uniform because material continues to
fall into the CND on short timescales.  Whether the further
infall of material from the CND toward the central black hole is
also transient and non-steady in nature is unclear from these studies.
However, there is some evidence that material may approach the central
region directly from outside of the CND via individual streamers.

{\it Acknowledgments} 
This work was supported in part by NSF Grant AST-21795 to the
University of California.  ALC is supported by an NSF Graduate Research
Fellowship. We thank Robert Zylka for permission to use the 1.2 mm image.

\newpage

\end{document}